\documentclass[aps,prb,twocolumn,amsmath,amssymb,superscriptaddress,floatfix,nofootinbib]{revtex4-2}

\usepackage{amssymb}
\usepackage{amsmath}
\usepackage{graphicx}
\usepackage{textcomp}
\usepackage{color}
\usepackage{xcolor}
\usepackage{comment}
\usepackage{amsfonts}
\usepackage{epsfig}
\usepackage{hyperref}
\usepackage{bm}
\usepackage[caption=false]{subfig} 
\usepackage{booktabs}
\usepackage{tabularx}
\usepackage{array}
\usepackage{soul}
\usepackage{wasysym}
\usepackage{feyn}
\usepackage{esint}
\usepackage{footmisc}
\usepackage{dcolumn}

\newcommand{\arctanh}[1]{\operatorname{arctan}}

\newcolumntype{M}[1]{>{\centering\arraybackslash}m{#1}}

\DeclareMathAlphabet\mathbfcal{OMS}{cmsy}{b}{n}

\begin{document}
\title{Half-metallic transport and spin-polarized tunneling through the van der Waals ferromagnet Fe${_4}$GeTe$_{2}$}

\author{Anita Halder}
\affiliation{School of Physics and CRANN, Trinity College, Dublin 2, Ireland}
\affiliation{Department of Physics, SRM University – AP, Amaravati 522 502, Andhra Pradesh, India}

\author{Declan Nell}
\affiliation{School of Physics and CRANN, Trinity College, Dublin 2, Ireland}

\author{Antik Sihi}
\affiliation{School of Physics and CRANN, Trinity College, Dublin 2, Ireland}

\author{Akash Bajaj}
\affiliation{School of Physics and CRANN, Trinity College, Dublin 2, Ireland}

\author{Stefano Sanvito}
\affiliation{School of Physics and CRANN, Trinity College, Dublin 2, Ireland}

\author{Andrea Droghetti}
\email[]{andrea.droghetti@tcd.ie}
\affiliation{School of Physics and CRANN, Trinity College, Dublin 2, Ireland}
\affiliation{Institute for Superconducting and other Innovative materials for devices, Italian National Research Council (CNR-SPIN), at G. d'Annunzio University, Chieti, Italy}

\begin{abstract}
The recent emergence of van der Waals (vdW) ferromagnets has opened new opportunities for designing spintronic devices. We theoretically investigate the coherent spin-dependent transport properties of the vdW ferromagnet Fe$_4$GeTe$_2$, by using density functional theory combined with the non-equilibrium Green's functions method. We find that the conductance in the direction perpendicular to the layers is half-metallic, namely it is entirely spin-polarized, as a result of the material's electronic structure. This characteristic persists from bulk to single layer, even under significant bias voltages, and it is little affected by spin-orbit coupling and electron correlation.  Motivated by this observation, we then investigate the tunnel magnetoresistance (TMR) effect in an magnetic tunnel junction, which comprises two Fe$_4$GeTe$_2$ layers separated by the vdW gap acting as insulating barrier. We predict a TMR ratio of almost 500\%, which can be further boosted by increasing the number of Fe$_4$GeTe$_2$ layers in the junction.
\end{abstract}
\maketitle

Magnetic tunnel junctions (MTJs), which consist of two metallic ferromagnets separated by a thin insulating barrier,
display the tunnel magnetoresistance (TMR) effect, that is a variation in the charge current when the magnetizations 
of the two ferromagnets change their relative alignments \cite{spintronics, parkin, JULLIERE1975, MTJ-1, Yuasa2004}. 
Recently, the discovery of magnetism in van der Waals (vdW) materials \cite{2dmagnets1,2dmagnets2}
has created new opportunities for realizing MTJs. A significant magnetoresistance was initially reported 
in devices with the insulating material CrI$_3$ sandwiched between graphite layers \cite{CrI3-MTJ1,CrI3-MTJ2}, 
while currently most studies focus on the Fe$_n$GeTe$_2$ (FGT) ($n$=3-5) family of vdW metallic ferromagnets 
\cite{FGT41st-dup}. 
In fact, various FGT-based MTJs incorporating h-BN \cite{hBN,hbn2}, graphite \cite{graphite}, MoS$_2$ \cite{MoS2}, 
InSe \cite{InSe}, GaSe \cite{GaSe}, WSe$_2$ \cite{WSe2}, or WS$_2$ \cite{WS} sandwiched between Fe$_3$GeTe$_2$ 
electrodes have been experimentally realized, recording a maximum TMR ratio of 300\% \cite{hbn2,hBN} and 
a current spin-polarization of up to 70\% \cite{WSe2} at low temperatures. At the same time, first-principles
calculations for similar systems \cite{tsymbol-mtj,Tsymbol_mtjhbn,mol-MTJ} predicted TMR ratios 
exceeding $1000\%$ or multiple nonvolatile resistance states.

Among the FGT compounds, Fe$_3$GeTe$_2$ has the lowest $T_\mathrm{C}$ (220 K) and requires sophisticated 
gating solutions to achieve room temperature ferromagnetism in few-layer samples \cite{FGT3}. 
Fe$_5$GeTe$_2$, despite having the highest $T_\mathrm{C}$ (310 K), is the most challenging to 
exfoliate \cite{FGT5}. Thus, Fe$_4$GeTe$_2$ (F4GT), with an intermediate $T_\mathrm{C}$ of 280 K, 
emerges as the most promising candidate for MTJs, due to its ease of exfoliation and preservation 
of ferromagnetism even in few-layer samples \cite{FGT41st-dup}. Furthermore, recent experimental 
investigations have demonstrated its potential for generating highly spin-polarized 
currents \cite{mukulprb}.

In this letter, we employ density functional theory (DFT) \cite{Kohn_nobel}, combined with the 
non-equilibrium Green's function (NEGF) technique \cite{Datta}, to investigate the spin-dependent 
coherent transport properties of F4GT from first-principles. Our findings reveal that the coherent 
transport perpendicular to the layers exhibits a half-metallic character, with one conductive spin channel
and an insulating one. This characteristic persists from bulk to monolayer, even under significant bias, 
so that a F4GT layer behaves as an almost perfect spin-filter. Furthermore, we show that the spin 
polarization is not degraded by the spin-orbit coupling (SOC) and electron correlation effects, unlike 
in other classes of half-metallic materials \cite{Pickett_2007,Mavropoulos_2004, quasi-rev}.
Finally, to further demonstrate the potential of the compound for spintronics, we predict a large TMR 
ratio in MTJs formed by stacking a few F4GT layers, with the vdW gap in between serving as an insulating 
barrier.

\begin{figure*}
\includegraphics[width=1.0\linewidth]{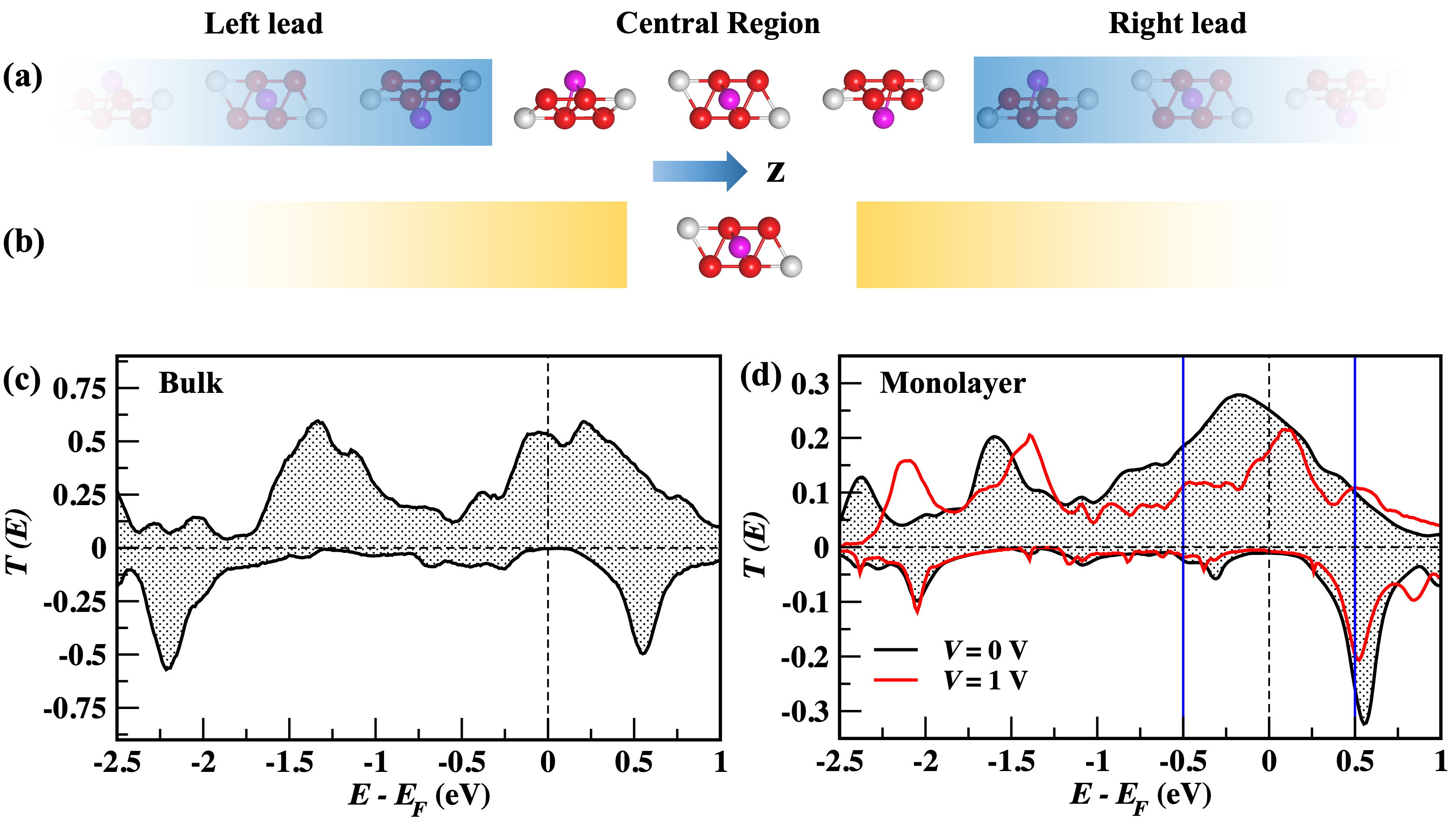}
\caption{Coherent transport through F4GT. (a) A device made of an infinite number of F4GT layers. 
Red, grey, and magenta spheres represent Fe, Te and Ge, respectively. (b) A device comprising a 
F4GT layer between model leads, represented as semi-infinite yellow rectangles. (c) Spin up 
(positive) and spin down (negative) transmission coefficient at zero-bias for the device in (a). 
(d) Spin up (positive) and spin down (negative) transmission coefficients at zero-bias and at 
a $V= 1$~V for the monolayer device in (b). 
The vertical blue lines delimit the bias window between $E_\mathrm{F} - eV/2$ and 
$E_\mathrm{F} + eV/2$ for $V=1$~V. }
\label{fig.struc}
\end{figure*}

The DFT-NEGF transport calculations are performed by using the {\sc Smeagol} code 
\cite{ro.ga.06, ivan_self_energies.ss.08,book1}, which interfaces the NEGF scheme with the {\sc Siesta} 
DFT package \cite{siesta}. We consider the Perdew-Burke-Ernzerhof generalized gradient approximation 
(GGA) \cite{PBE} for the exchange-correlation functional in all calculations, unless stated otherwise. 
The computational details are provided in Section S1 of the supplementary information (SI). 
The studied systems are shown in Figs. \ref{fig.struc} (a) and (b) and consist of a central 
scattering region and two semi-infinite leads. A finite bias voltage, $V$, is applied 
across the central region by shifting the chemical potentials of the leads as $\mu_\mathrm{L/R} = 
E_\mathrm{F} \pm eV/2$, 
where $E_\mathrm{F}$ is the Fermi energy and $e$ the electron charge. Both zero- and finite-bias 
calculations are performed self-consistently. 

We initially assume a two-spin-fluid picture \cite{Mott} for coherent charge transport and 
perform spin-collinear calculations, following common practice in the study of MTJs \cite{JULLIERE1975}. 
Under this assumption, the two spin channels conduct in parallel without mixing, and the charge current 
for spin $\sigma$ $(\sigma=\uparrow,\downarrow)$ is defined as \cite{Datta}
\begin{equation}\label{eqn: LB}
    I^{\sigma} = \frac{e}{h} \int dE \big[f_L(E )-f_R(E)\big] T^{\sigma}(E, V),
\end{equation}
where $h$ is Planck's constant, $f_\mathrm{L(R)}(E)= [1+e^{\beta(E-\mu_\mathrm{L(R)})}]^{-1}$ the Fermi 
function of the left (right) lead, and $\beta$ is the inverse temperature. The spin-, energy- and bias-dependent transmission coefficient,
$T^{\sigma}(E,V)$, is calculated through the Fisher-Lee formula \cite{PhysRevB.23.6851}. According to 
Eq. (\ref{eqn: LB}), the transport is determined by the coherent transmission of spin up and down electrons 
from one lead, through the central region, to the other lead. The transmission coefficient depends
on $V$ because the electronic states may shift in energy under the applied bias (see, for 
instance, Refs. \cite{IvanFeMgO,mol-MTJ}). 
Notably, $I^\sigma$ in Eq. (\ref{eqn: LB}) is approximately equal to the area under the transmission 
coefficient-vs-energy curve inside the energy interval [$E_\mathrm{F} - eV/2$, $E_\mathrm{F} + eV/2$], 
known as bias window. 
In the linear response limit and at zero temperature, the expansion of Eq. (\ref{eqn: LB}) returns the conductance of each spin channel through the Landauer-B\"uttiker formula,
$ G^{\sigma} = G_0 T^{\sigma}(E_\mathrm{F}, V=0)$ \cite{La.57,Bu.86,Bu.88}, with $G_0=\frac{e^2}{h}$ 
denoting the quantum of conductance. In the following, the dependence of the transmission coefficient 
on $V$ will no longer be explicitly indicated to keep the notation concise.
\begin{figure*}
\includegraphics[width=1.0\linewidth]{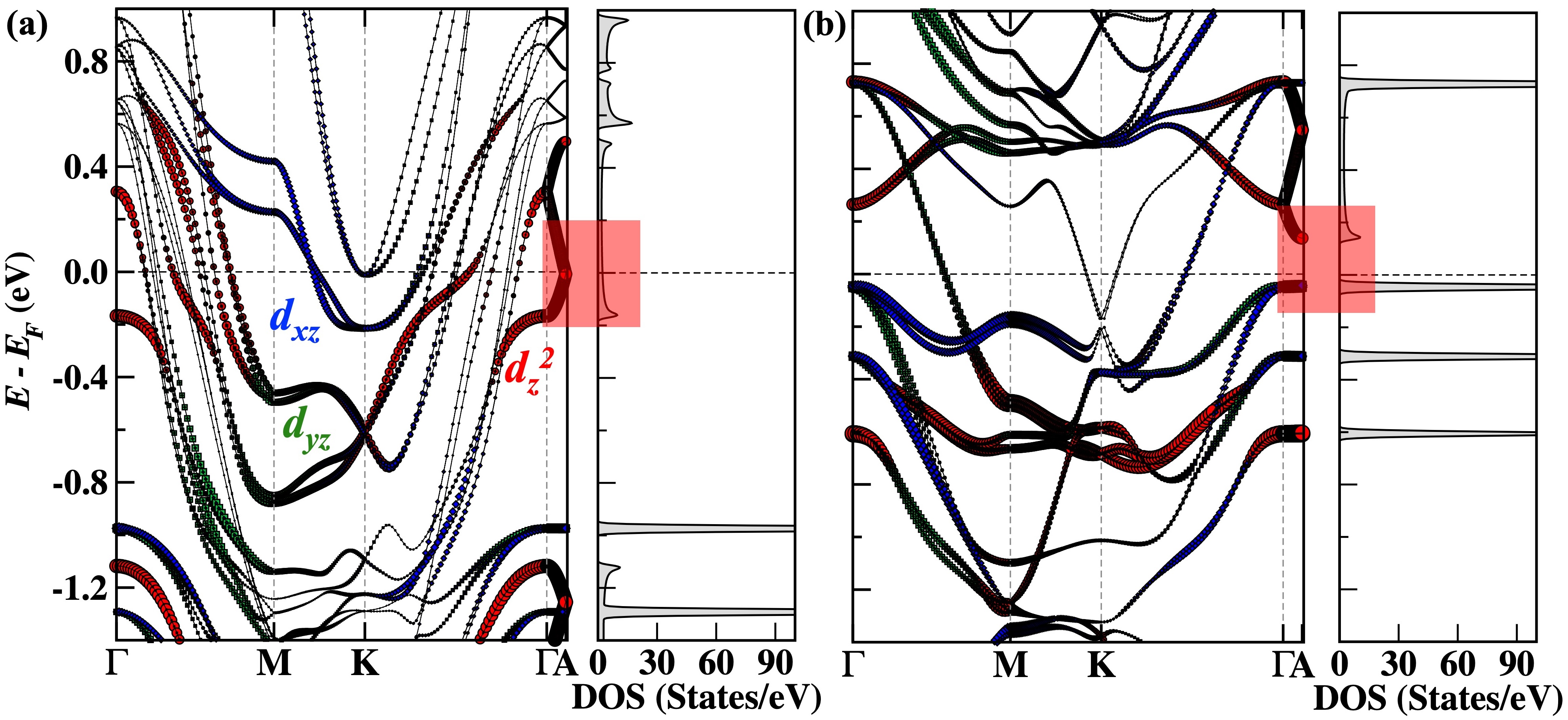}
\caption{Band structure of bulk F4GT: (a) spin up, (b) spin down. The width and colour of the bands 
indicate the orbital character. Bands with Fe $3d_{xz}$, $3d_{yz}$ and $3d_{z^2}$ orbital character 
are blue, green and red, respectively. On the right-hand side of each band structure, we show the 
density of the states with momentum $\hbar k_z>0$ along the $\Gamma$-$A$ direction in the Brillouin 
zone. The energy region around $E_\mathrm{F}$ is highlighted in red.  
}
\label{bands}
\end{figure*}

We start by calculating the transport properties of a device that consists of an infinite number of 
F4GT layers with ABC stacking \cite{FGT41st-dup}, as shown in Fig. \ref{fig.struc}(a). We consider the 
transport perpendicular to the layers, that is along the $z$ direction, with periodic boundary conditions 
in the $xy$ plane. More details can be found in Section S1 of the SI. The zero bias spin-resolved 
transmission coefficient is plotted in Fig. \ref{fig.struc}(c), showing a remarkable half-metallic behaviour. 
In fact, $T^{\uparrow}(E)$ exhibits a prominent peak, whereas $T^{\downarrow}(E)$ is gapped around 
$E_\mathrm{F}$. According to the Landauer-B\"uttiker formula, the linear-response conductance for the 
spin-down channel is negligible compared to the one for spin up. The spin-polarization, defined 
as $\textnormal{SP} = \frac{G^{\uparrow}-G^{\downarrow}}{G^{\uparrow}+G^{\downarrow}}$, is as high 
as 0.99.

To obtain a better understanding of the half-metallic transport 
behaviour, in Fig.~\ref{bands} we plot the spin-resolved band structure 
of bulk F4GT, where the blue, green and red bands have predominant amplitude 
over the Fe $3d_{xz}$, $3d_{yz}$ and $3d_{z^2}$ orbitals, respectively. Our 
results, which agree well with a published work \cite{tsymbol-mtj}, 
reveal that the band structure is not half-metallic with both majority and 
minority bands cutting $E_\mathrm{F}$ at several points in the Brillouin zone. 
However, the situation is different when we restrict the analysis only to the 
$\Gamma$-$A$ direction, where the momentum $\hbar k_z>0$ is perpendicular to 
the layer. In the majority channel, there is a band with $d_{z^2}$ character 
crossing $E_\mathrm{F}$, 
whereas the minority 
channel has a band gap with minimum $E_g \sim 0.2$~eV at $A$, and dispersionless 
valence bands. As a consequence, only majority Bloch states originating from 
the inter-layer hybridization of the $3d_{z^2}$ orbitals can carry current in 
the perpendicular direction, giving rise to perfect spin polarization in 
the linear response limit. The density of the states (DOS) along the 
$\Gamma$-$A$ direction (shown on the right hand side of the band structure) 
is typical of a half-metal. 

The half-metallic behaviour of F4GT is preserved when moving away from bulk, 
down to the extreme monolayer limit. This is seen in the device of 
Fig.~\ref{fig.struc}(b), featuring one F4GT layer stacked between two ``model'' 
leads, mimicking the wide-band approximation \cite{haug1996quantum}. The 
leads' impact on the central region's electronic structure is merely to 
broaden and smooth the DOS (shown in Fig.~S3 of the SI) without introducing 
any interface states or proximity effects. In this way, the transport properties 
can be directly associated to the intrinsic electronic structure of the F4GT 
layer and, in particular, to the energy position of the Fe $3d_{z^2}$ states 
with respect to $E_\mathrm{F}$.
The transmission coefficient for such a device is depicted in 
Fig.~\ref{fig.struc}(d). This is qualitatively similar to one of bulk F4GT, 
displaying a prominent peak in the spin up channel around $E_\mathrm{F}$ and a gap in the spin down one. $G^\uparrow$ is about 0.25$G_0$, 
while $G^\downarrow$ is several orders of magnitude smaller. As such, the monolayer 
in our model device effectively acts as an almost ideal spin-filter.

The spin down transmission coefficient exhibits a significant asymmetry. 
The peak associated with conduction states around $E-E_\mathrm{F}\sim 0.5$ is much higher than the one corresponding to valence states at $E-E_\mathrm{F}\sim -0.3$. This difference arises from the distinct localization of these states, as detailed in Section S5 of the SI, and, in practical terms, implies the persistence of nearly perfect spin-polarization even under p-doping conditions, which are common in F4GT owing to Te vacancies.

\begin{figure}[h]
\includegraphics[width=1.0\linewidth]{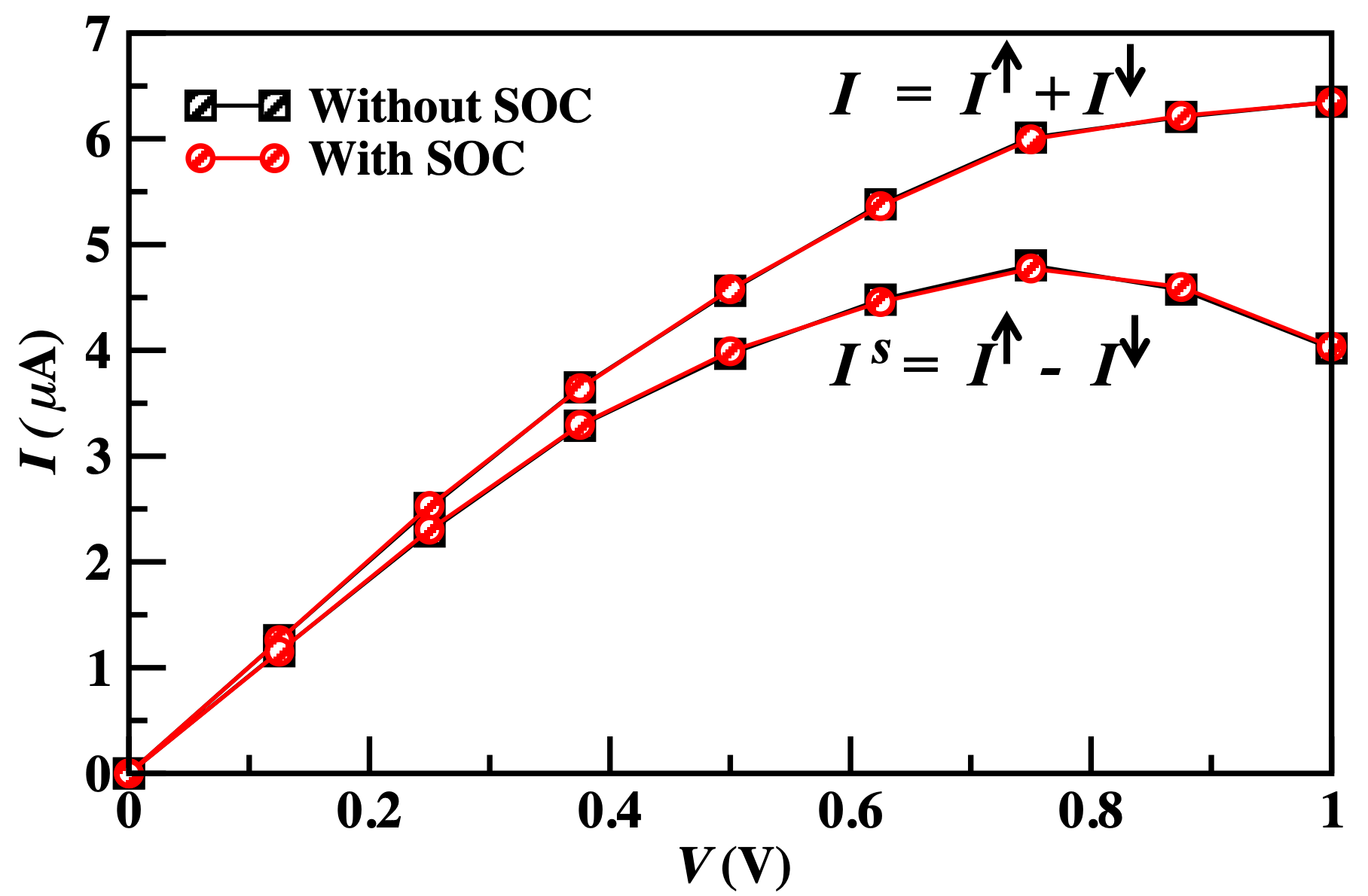}
\caption{Results of the finite-bias calculations for the monolayer device. 
Charge and spin currents, $I$ and $I^s$, as a function of bias voltage, $V$. 
Black (red) points are the results obtained without (with) SOC.}
\label{fig.bias3}
\end{figure}

The study of the F4GT-monolayer device can be further extended beyond the 
linear-response limit by performing finite bias calculations. 
The electronic structure is found to change with the bias, $V$ as explained in Section~S6 
of the SI. However, the transmission coefficient [red curve in Fig. \ref{fig.struc}(d)]
remains half-metallic 
with a spin-down gap at $E_\mathrm{F}$, similar to what was found for $V=0$ V. 
The charge and spin currents, respectively defined as 
$I = I^\uparrow + I^\downarrow$ and $I^s = I^\uparrow - I^\downarrow$, are 
plotted in Fig.~\ref{fig.bias3} as a function of $V$. 
These curves are understood by recalling that $I^{\uparrow(\downarrow)}$ 
is approximately equal to the area under the spin up (down) transmission curve 
inside the bias window [see Eq.~(\ref{eqn: LB})], which is delimited by the blue 
bars in Fig.~\ref{fig.struc}(d).
At low biases ($V \lesssim 0.3$ V), $I^\uparrow$ dominates while $I^\downarrow$ 
is negligible because of the half-metallic character of the transmission coefficient. 
Thus, $I$ (solid curve) and $I^s$ (dotted curve) are identical, and the current 
spin polarization, $I^s/I$, is about 1. In contrast, at high biases 
($V \gtrsim 0.6$ V), 
$I^\downarrow$ starts increasing with $V$ as the spin down gap's edges enter the 
bias window [see Fig.~\ref{fig.struc}(d)]. The electrons from the minority valence 
and conduction states then contribute to the transport in parallel with the majority
electrons, reducing the spin current. Hence, the system does not show 
half-metallic conduction anymore. Nonetheless, we note that the spin-polarization 
remains as large as $\sim0.7$ at $V=1$ V. Thus, a F4GT monolayer acts as an 
effective spin-filter even up to high biases.

We now analyse whether the predicted behaviour is modified by some 
potentially important electronic effects that have been neglected up to this point.
Firstly, we introduce SOC in the calculations \cite{SOC_onsite}. This provides a
mechanism for spin mixing, invalidating the two-spin-fluid picture. SOC is 
known to degrade the perfect spin-polarization in many materials otherwise predicted 
to be half-metals \cite{Mavropoulos_2004, Pickett_2007}. In F4GT, its effect on 
the electronic properties is significant, 
as confirmed by inspecting the bulk band
structure with SOC [Fig.~S1 in the SI]. Nevertheless, we find that, even with SOC, there is only one band, with well defined spin character, crossing 
the Fermi level along the $\Gamma$-$A$ direction. Hence, we conclude that 
the transport properties of bulk F4GT are not modified by SOC. 

For the monolayer case, we extend the calculations with SOC to finite bias. 
Now the spin-resolved currents are no longer defined, but we use the so-called ``bond
current'' approach \cite{Todorov_2002} to derive a general definition of the spin current,
$I_s$, valid also in presence of SOC \cite{PhysRevB.105.024409, bajaj2024intrinsic}. 
The results are presented as red circles in Fig.~\ref{fig.bias3}. They appear indistinguishable from those previously obtained without 
SOC (black squares), confirming that spin-mixing is negligible in the 
transport through our systems, and the predictions based on the two-spin-fluid picture, 
are reliable.

Next, we investigate the effect of electron correlations beyond the GGA. In general, 
electron correlations impact transport through ferromagnetic metallic layers by inducing 
an energy shift of the conductive electronic states \cite{liviu_Cu_Co_dmft,andrea_Cu_co}.
Furthermore, in half-metals, dynamical correlations may also give rise to non-quasiparticle 
peaks in the insulating spin channel \cite{quasi-rev}, thus quenching the perfect spin-polarization. In the case of the FGT compounds, the importance of electron correlations 
remains debated. Yet, experimental observations suggest that there exists competition 
between itinerant and localized electrons \cite{nonstonerFGT}. Similarly, 
theoretical studies \cite{bsanyal} report that dynamical correlation included within 
Dynamical Mean-Field Theory (DMFT) \cite{dmft1,dmft2} is essential to accurately
describe the magnetic properties. FGT compounds have therefore been regarded as 
moderately correlated materials.

Here, we carry out calculations by using DFT+U and DFT+DMFT. In DFT+U, an effective 
Hubbard-like $U$ interaction for the Fe $3d$ orbitals is added to the GGA exchange-correlation
functional and is treated at the static mean-field level \cite{an.za.91,li.an.95,du.bo.98}. 
In contrast, DFT+DMFT accounts also for dynamical correlations (albeit local in space) via 
an energy-dependent self-energy \cite{dmft1,dmft2}. Hence, a comparison of the DFT+U and 
DFT+DMFT results gives some insights into the relative importance of static and 
dynamical correlations. Here, we focus only on the zero-bias limit and restore the 
two-spin-fluid picture, which we have just shown to be appropriate for our system. We employ 
the implementation of the methods described in Refs. \cite{andrea_Cu_co,andrea_ivan_projection,sigma2} with the computational details in the SI. 

\begin{figure}
\includegraphics[width=1.0\linewidth]{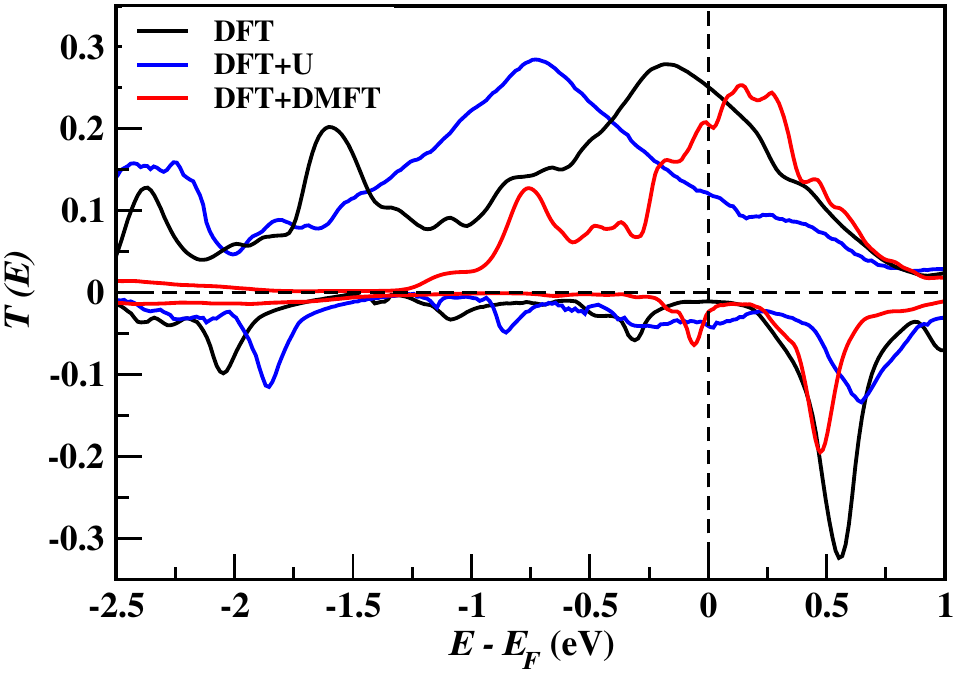}
\caption{Zero-bias transmission coefficient calculated by using DFT (black curve), DFT+U 
(blue curve) and DFT+DMFT (red curve).}
\label{fig.struc4}
\end{figure}  
\begin{figure*}
\includegraphics[width=1.0\linewidth]{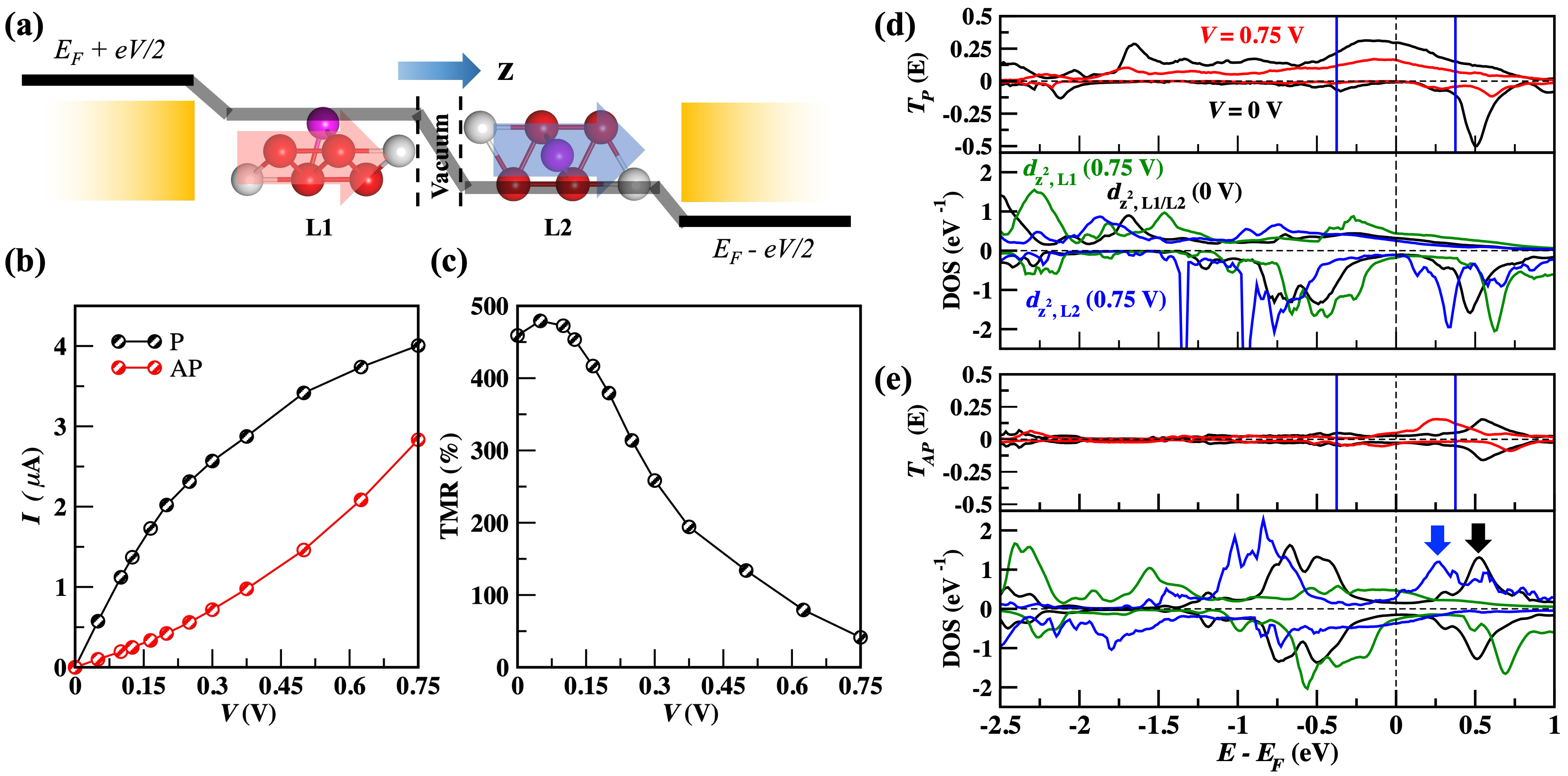}
\caption{Results for the F4GT-based MTJ. (a) The MTJ consists of two F4GT layers, denoted 
as $L1$ and $L2$, sandwiched between model leads, represented as semi-infinite yellow 
rectangles. The transport direction is along $z$ Cartesian axis. The magnetization vectors 
of the two F4GT layers in the P configuration are pictured as thick red and blue arrows for 
L1 and L2, respectively. The electrostatic potential in the central region is also shown
schematically as a black thick line. It drops linearly across the vdW gap, while it remains 
nearly constant inside the F4GT layers. 
(b) The current-voltage characteristic curve for the P and AP configurations. 
(c) TMR ratio as a function of bias voltage. 
(d) The transmission coefficient (upper panel) and the DOS projected over Fe $3d_z^2$ orbitals 
of L1 and L2 (lower panel) for the P configuration at zero bias and at $V=0.745$ V. 
(e) Same as (d) for the AP configuration. Note that the spin up and spin down DOS do not look identical in the AP configuration because the system is not exactly inversion-symmetric with respect to the center of the device. 
The black (blue) arrow indicates the position of 
the spin up conduction states of L2 at zero bias ($V=0.745$ V).}
\label{fig.struc5}
\end{figure*}

The transmission coefficients of the monolayer device calculated by DFT+U and 
DFT+DMFT are presented in Fig.~\ref{fig.struc4}, while the DOS are 
provided in Section S7 of the SI. Static correlations, as described by DFT+U, enhance 
the spin splitting of the Fe $3d_{z^2}$ states compared to DFT. As a result, in the 
spin up channel, the transmission's main peak moves from $E_\mathrm{F}$ to lower 
energies by about 0.5 eV. Therefore, the linear-response conductance, 
$G^\uparrow \propto T^\uparrow(E_\mathrm{F})$, is reduced by more than half with 
respect to the DFT value. Conversely, in the minority channel, the center of the 
transport gap shifts away from $E_\mathrm{F}$ towards higher energies, so that the 
valence states cross $E_\mathrm{F}$. 
Hence, the half-metallic character is suppressed. Overall, DFT+U 
reduces the linear-response spin-polarization to $\sim0.5$.

The inclusion of dynamical correlation by means of DMFT redistributes the Fe $3d$ 
states in energy, counterbalancing the effect of static correlation, and reducing 
the spin splitting. For spin up, the main transmission peak is narrowed, but, once 
again, centered near $E_\mathrm{F}$. For spin down, the transmission remains 
insulating, although the size of transport gap is reduced with respect to the DFT one, and the valence states get pinned at $E_\mathrm{F}$. We find no 
non-quasi-particle peaks, unlike in other half-metallic materials. The change of 
the transmission coefficient from DFT(+U) to DFT+DMFT calculations can be ascribed 
uniquely to the energy shift and the finite lifetime of the $3d$ quasi-particle 
states (see Section S7 in the SI). The reduction of the spin down transmission's gap can, in principle, 
be detrimental to the half-metallic behaviour, however in practice, the spin down 
valence states 
are so minimally conductive in comparison with the spin up states  
that the spin-polarization remains almost unchanged.

Overall, these calculations indicate that, although electron correlations beyond DFT are
important in F4GT, the combined effect of static and dynamical contributions preserves 
the almost perfect spin-filter character predicted by DFT. 

The nearly half-metallic transport of F4GT can eventually be exploited in 
a MTJ. This possibility is explored by considering the idealized device of 
Fig.~\ref{fig.struc5}(a). The central region, attached to the same model leads used 
before, comprises of two F4GT layers (L1 and L2), separated by the vdW gap that serves 
as insulating barrier. The device can be set in two configurations, with the 
magnetization vectors of the two F4GT layers being either parallel (P) or antiparallel 
(AP) to each other. The calculations are carried out by using spin-collinear DFT, which
captures the key transport features, as explained before. The charge current as a 
function of the applied bias voltage for the two configurations is displayed in 
Fig.~\ref{fig.struc5}(b). At low bias, the P current, $I_{\textnormal{P}}$, is 
significantly larger than the AP current, $I_{\textnormal{AP}}$. In contrast, with 
increasing bias, $I_{\textnormal{P}}$ tends to saturate, while $I_{\textnormal{AP}}$ 
sharply increases. 
As a result, the TMR ratio, defined as $(I_\textnormal{P} - I_{\textnormal{AP}}) / I_{\textnormal{AP}}$, is as large as $460\%$ at low bias ($V<0.15$ V), and then drops 
with $V$, becoming about $50\%$ at 0.75~V. 
At zero bias, the TMR is understood through the standard Julliere's phenomenological description \cite{JULLIERE1975}. We assume transport from left to right so that the left 
F4GT layer (L1) filters spin up electrons, which are then detected by the right layer 
(L2). In the P configuration, since L2 is metallic in the spin up channel, spin up 
electrons are transmitted through. In contrast, in the AP configuration, the L2 spin 
up channel turns insulating, thus that transport is greatly suppressed. As a result, 
the TMR is large. 
Quantitatively, the effect is analyzed in terms of transmission coefficients of 
Figs.~\ref{fig.struc5}(d) and \ref{fig.struc5}(e) for the P and AP configurations, 
respectively. $T^\sigma_\mathrm{P}(E)$ appears similar to its counterpart for
the monolayer device, and the conductance $G^\uparrow$ is as large as $\sim0.25$ $G_0$. 
In contrast, $T^\sigma_\mathrm{AP}(E)$, which is identical in both spin channels, 
is approximately given by the convolution of the spin up (metallic-like) and spin down (insulating-like) transmissions of the monolayer, as expected based on the 
standard model of MTJs \cite{SanvitoReview,Rungger2020}. As such, it nearly vanishes 
at $E_\mathrm{F}$. 

At finite bias, the P and AP currents, and therefore the TMR, depend on change of 
the energy alignment of the L1 and L2 Fe $3d_{z^2}$ states. 
This is understood by noting that the electrostatic potential predominantly drops 
across the vdW gap between the two F4GT layers, as schematically drawn in 
Fig.~\ref{fig.struc5}(a). Consequently, the states in L1 (L2) are pinned to the left 
(right) lead and experience an upwards (downwards) energy shift with increasing $V$. 
In the P configuration [Fig. \ref{fig.struc5}(d)], the Fe $3d_{z^2}$ DOS of L1 and 
L2 becomes misaligned with $V$, leading to a reduced electronic overlap through the 
vdW barrier, and to a partial suppression of the transmission coefficient. 
In the AP configuration [Fig. \ref{fig.struc5}(e)], we see a somewhat opposite behaviour. 
The spin up channel of L2 is insulating preventing transport at low bias. Yet, with 
increasing $V$, its conduction states [indicated by the arrows in bottom panel of Fig. \ref{fig.struc5}(e)] move down in energy until they eventually enter the bias window.
When that happens, the electrons filtered by L1 can be transmitted though L2, leading 
to a sharp increase in the AP current with bias, and therefore to 
a drop in the TMR ratio.

Interestingly, a previous quantum transport study \cite{tsymbolprl} recently predicted 
a TMR of only $\sim24\%$ at zero bias in an MTJ made of two F4GT layers ($1\Bar{1}0$) 
separated by the vdW gap. However, we note that in that case the transport was in-plane, 
while, as shown here, the half metallic conductance is characteristic only of the 
perpendicular direction. In this perpendicular case, the TMR at low bias can be 
further enhanced just by increasing the number of F4GT layers acting as spin-filters. 
For example, calculations for a three-layer device, presented in Section S8 of the SI, give a 
huge TMR ratio, exceeding 1200\% at zero bias. This value is comparable to the one 
predicted in Fe(001)/MgO MTJs \cite{bu.zh.01} used in technological applications. 

In practice, operating F4GT-based MTJs in experiments requires the capability of 
switching the magnetization of a layer independently from that of the others. This 
can be achieved, for example, by substituting some of the perfect F4GT layers with 
slightly off-stoichiometric compounds, such as Fe$_{4-x}$GeTe$_2$ \cite{off-stoichio}, 
characterized by a different coercive field. Alternatively, one may place the 
spin-filter and detector layers in contact with leads made of different heavy metals, 
thus tuning their relative magnetic anisotropy by proximity. 

In summary, our first-principles calculations have revealed the robust spin-filtering
capability of the vdW ferromagnet F4GT along the out-of-plane direction, and we have 
then explored the potential of this effect for MTJ applications. 
The predicted zero-bias TMR ratio is about 460\% for a two-layer MTJ and rises with the 
number of F4GT layers, eventually exceeding 1000\%. This, combined with the possibility
of fabricating devices by mechanical exfoliation, opens up the possibility of using 
vdW ferromagnets for high-performing spintronics devices.



\section*{Acknowledgements}
A.H. was supported by European Commission through the Marie Skłodowska-Curie individual fellowship VOLTEMAG-101065605.
A.D. and A.S. were supported by Science Foundation Ireland (SFI) and the Royal Society through the University Research Fellowship URF/R1/191769.
D.N. was supported by the Irish Research Council (Grant No. GOIPG/2021/1468). 
A.B. and S.S. were supported by SFI (19/EPSRC/3605) and of the Engineering and Physical Sciences Research Council (EP/S030263/1).
Computational resources were provided by Trinity College Dublin Research IT and the Irish Center for High-End Computing (ICHEC). 

\section*{Supplementary Information}
\setcounter{figure}{0}
\renewcommand{\thesection}{S\arabic{section}}
\renewcommand{\thefigure}{S\arabic{figure}}
\renewcommand{\thetable}{S\arabic{table}} 
\section{Computational details}
\subsection{F4GT structures and geometry optimizations}
Bulk F4GT possesses a rhombohedral crystal structure with a {\it R-3m} space group, and experimental lattice parameters, $a = 4.04$~\AA~and $c = 29.08$~\AA \cite{tsymbol-mtj, FGT41st-dup}. The primitive unit cell contains three layers separated along the $c$-axis by a van der Waals (vdW) gap. There are four Fe atoms, Fe1, Fe1', Fe2, and Fe2', in each layer (see the inset of Fig. \ref{SI2}). They occupy two inequivalent Wyckoff positions forming two pairs of Fe-Fe dumbbells directly bonded to the Te atoms. Fe1 and Fe2 are equivalent to Fe1' and Fe2'. The Ge atoms are positioned off the plane defined by the Fe-Te network. The layers are stacked in an ABC configuration along the $c$-axis, as shown in Fig. 1(a) in the paper.  We assume a Cartesian frame of reference such that the $c$ axis is parallel to the Cartesian $z$ axis. 
For an F4GT monolayer, we adopt the lattice constant, $a = 3.92$ \AA, from Ref. \onlinecite{FGT4-monolayer}, which is slightly shorter than the bulk one.

All atomic positions are optimized by using DFT as implemented in the Vienna {\it Ab-initio} Simulation Package ({\sc VASP})~\cite{VASP}. We employ the Perdew-Burke-Ernzerhof (PBE) generalized gradient approximation (GGA) \cite{PBE} for the exchange-correlation functional with the vdW corrections included within the DFT-D3 scheme \cite{DFTD3}. The calculations are spin-polarized.
A $\Gamma$-centered $\mathbf{k}$-mesh of 30 $\times$ 30 $\times$ 1 and convergence criteria of 10$^{-7}$ eV for total energy are employed. Structures are relaxed until all forces on the atoms are smaller than 10$^{-3}$ eV/\AA.
The Gaussian smearing method is used with a kinetic-energy cutoff of 600 eV.

The monolayer, the two-layer and three-layer structures used in the magnetic tunnel junctions (MTJ) are optimized considering a supercell with $30$~\AA~of vacuum separating periodic images along the $z$ direction. The calculated vdW gap in the two-layer system is $3.3$~\AA~similar to the one reported in Ref. \cite{FGT41st-dup}.

\subsection{Band structure calculations}

Spin-collinear DFT calculations are performed with the PBE GGA exchange correlation functional\cite{PBE} using the PSML (pseudopotential markup language) compatible version of the {\sc Siesta} DFT package\cite{siesta, siestapsml}. Core electrons are treated using norm-conserving Troullier-Martins pseudopotentials\cite{pseudopotential_1, pseudopotential_2}. The $spd$ valence electrons are expanded using the numerical atomic orbital basis set of double-$\zeta$ quality, while additional polarization functions are incorporated into the 4$s$ Fe orbitals\cite{artacho1999linear,junquera2001numerical}. The cutoff radii of the basis orbitals are taken from Ref. 
\onlinecite{rivero2015systematic}. The pseudopotentials and the basis set are validated by comparing the band structure with the one obtained from {\sc VASP} calculations. 


A $12\times12\times8$ $\Gamma$-point centered Monkhorst-Pack \textbf{k}-mesh is used. The plane-wave cutoff corresponding to the resolution of the real-space density grid is set to 600 Ry. 
The orbital-resolved band structure is plotted by means of the \texttt{fat} and \texttt{eigfat2plot} utilities available within {\sc Siesta}. 
The corresponding projected density of states (PDOS) is calculated considering 60 $\mathbf{k}$-points along the $\Gamma-A$ direction and a Lorentzian broadening equal to 0.01 eV.  

The calculations including spin-orbit coupling (SOC) are carried out by using a locally modified version of {\sc Siesta}. The on-site approximation of Ref. [\onlinecite{SOC_onsite}] is assumed for the spin-orbit matrix elements. All computational parameters are the same as in calculations without SOC. 

\subsection{Quantum transport calculations}

The quantum transport calculations are performed by using the NEGF method as implemented in the {\sc Smeagol} code \cite{ro.ga.06, book1} interfaced with {\sc Siesta}. The basis set, pseudopotentials and exchange-correlation functional are the same as in the band structure calculations.

The leads are treated via self-energies which are calculated using a singular value decomposition-based algorithm\cite{ivan_self_energies.ss.08}.
Both zero and finite bias calculations are performed self-consistently. The electrostatic Hartree potential, $V_\mathrm{H}$, inside the central region is obtained by solving the Poisson equation. The difference, $\Delta V_\mathrm{H}$, between the Hartree potential at finite and zero bias gives the voltage drop across the central region.

The density matrix of the central region is calculated by splitting the integration of the lesser Green's function into the so-called equilibrium and non-equilibrium components \cite{ro.ga.06}. 
The equilibrium component is obtained by performing the integration over a semicircular contour in the complex energy plane \cite{ro.ga.06}. 16 poles are used in the Fermi distribution, and 16 energy points are used along both the semicircle and the imaginary line that form that contour. The non-equilibrium component is calculated by performing the integration over the real energy axis using 100 energy points inside the bias window.

The self-consistent density matrix is calculated with a 20 $\times$ 20 $\Gamma$-point centered Monkhorst-Pack \textbf{k}-mesh. The converged density matrix is then read as an input for non-self consistent calculations using a 50 $\times$ 50  (100 $\times$ 100) \textbf{k}-mesh to obtain the transmission coefficient and the DOS of bulk F4GT (of the monolayer device and of the MTJs). 

The bond current approach, described in Refs. [\onlinecite{PhysRevB.105.024409, bajaj2024intrinsic}], is employed to calculate the charge and spin currents in the presence of SOC. In spin-collinear calculations, it was verified that the results obtained via the bond current approach are identical within two significant digits to the current obtained from Eq. (1) in the paper.

\subsection{DFT+DMFT calculations}
DFT+DMFT calculations for the monolayer device are carried out with {\sc Smeagol} by using the implementations in Refs. [\onlinecite{sigma2}] and [\onlinecite{andrea_Cu_co}]. The correlated subsystem spanned by the Fe $3d$ orbitals is downfolded from the device central region by means of the scheme in Ref. [\onlinecite{andrea_ivan_projection}]. General orbital-dependent Coulomb interaction parameters for the $3d$ orbitals within each Fe atom are considered. These are expressed in terms of Slater integrals $F^0$, $F^2$ and $F^4$ (Ref. [\onlinecite{im.fu.98}]). The ratio $F^4/F^2$ is assumed to correspond to the atomic value $\approx 0.625$ (Ref. [\onlinecite{an.gu.91}]). 
The average $U$ and $J$ interaction parameters are given through the relations $U=F^0$ and $J=(F^2+F^4)/14$. 

Second-order perturbation theory in the electron-electron interaction is employed as an impurity solver for DMFT \cite{sigma2} allowing for the fast evaluation of the self-energy directly on the real energy axis with no need for any analytic continuation schemes. The first order contribution accounts for static (energy-independent) mean-field corrections to the DFT GGA Kohn-Sham Hamiltonian and is approximated with the $U$-potential by Dudarev {\it et al.} \cite{dudarev} For each Fe $3d$ orbital, $\lambda$, this has the form, 
\begin{equation}\label{eqn: dudarev}
    V_{U, \lambda}^{\sigma} = (U-J)(\frac{1}{2} - n_{\lambda}^{\sigma}),
\end{equation} where $\sigma$ denotes the spin and $n_{\lambda}^{\sigma}$ is the occupation. In practice, the use of only this first order correction coincides with performing DFT+U calculations. The potential $U$ in Eq. (\ref{eqn: dudarev}) is positive (negative) for orbitals with less (more) than half-filling occupations $n_{\lambda}^{\sigma}$.

The second order contribution introduces dynamic correlation. The local Green's function is calculated by summing the retarded Green's function over a $20 \times 20$ $\Gamma$-point centered Monkhorst-Pack $\mathbf{k}$-mesh. The many-body self-energy is computed on an energy grid comprising 3000 points extending from -20 to 10 eV. DFT+U calculations are performed evaluating the charge density self-consistently. DFT+DMFT calculations are not, i.e., the DMFT self-energy is obtained by solving the self-consistent DMFT equations, but the resulting charge density is not used to re-start the DFT part of the calculation.

\begin{figure}[h]
\includegraphics[width=1.0\linewidth]{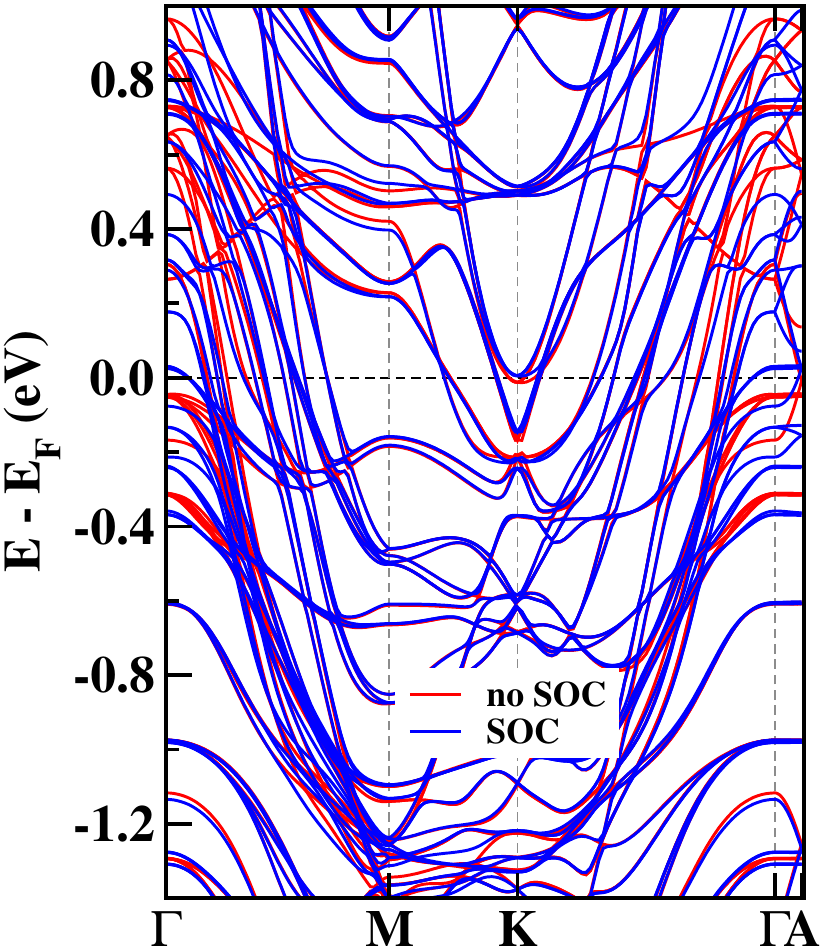}
\caption{Band structure of bulk F4GT with (blue) and without SOC (red).}
\label{bssoc}
\end{figure} 
\section{Band structure with SOC}

Fig. \ref{bssoc} displays the band structure of bulk F4GT calculated with and without SOC (blue and red curves, respectively). The SOC has a quite strong effect around the $\Gamma$ point, where it induces a large ($\sim 100$ meV) splitting of several otherwise degenerate bands. Moving along the $\Gamma$-A direction, relevant for perpendicular transport, we observe that there is only one dispersive band crossing the Fermi level, as explained in the paper. This band is weakly affected by the SOC meaning that it preserves a well defined spin character, and therefore the transport is expected to remain half-metallic.



\section{Complex band structure of bulk F4GT}

\begin{figure*}
\includegraphics[width=0.75\linewidth]{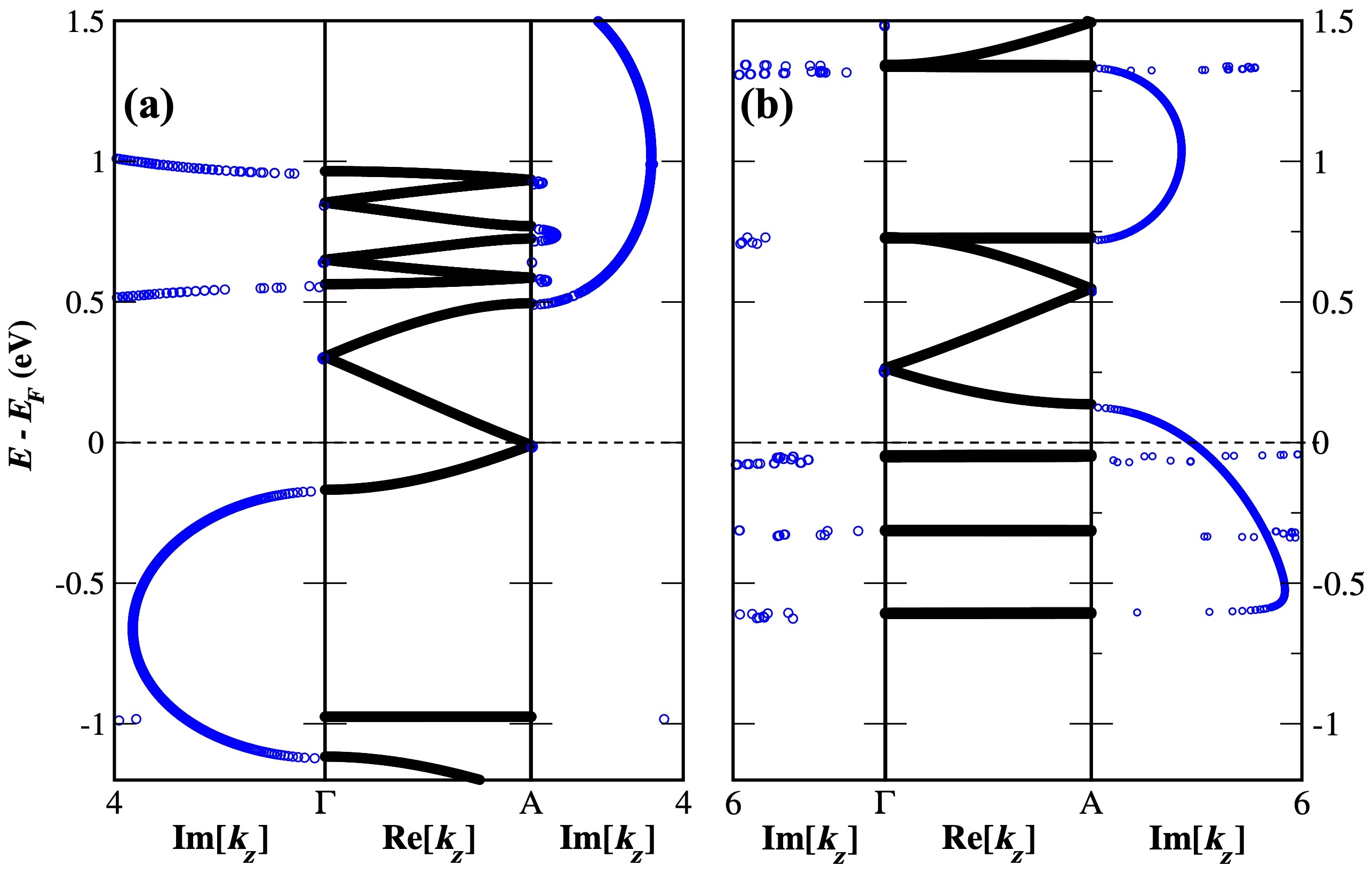}
\caption{Complex band structure of F4GT for (a) spin up  and (b) spin down.}\label{complex_bs_smeagol}
\end{figure*}

The half-metallic character of bulk F4GT for perpendicular transport can be clearly visualized from the complex band structure \cite{PhysRev.115.809, Heine_1963,PhysRevB.73.035128, Reuter_2017} which is obtained as an output of the algorithm used to compute the leads' self-energies\cite{ivan_self_energies.ss.08} in the {\sc Smeagol} code (an alternative and equally valid algorithm is described in Ref. [\onlinecite{Bosoni_2022}]). The complex band structure generalizes the conventional band structure by considering wave-vectors with complex components and, therefore, describes the bulk-propagating states as well as the evanescent states that decay across the F4GT vdW gap.

Fig. (\ref{complex_bs_smeagol}) displays the calculated band structure (black lines) with real wave-vector $k_z$ along the $\Gamma$-A direction in the Brillouin zone, and the complex band structure (blue lines) with $\textnormal{Im}[k_z]\neq 0$ at the $\Gamma$ point and at the $A$ point. The (a) and (b) panels are respectively for spin up and spin down. The real band structure is identical to that presented in Fig. \ref{bssoc} and in Fig. 2 of the paper. For spin up, there are only real bands, whereas, for spin down, there is a band gap at the Fermi level. The lowest spin down conduction band terminates at the A point at an energy $E-E_\mathrm{F}\sim0.15$ eV, and then continues as a complex band. Thus, while the transport of spin up electrons is due to bulk-propagating states, the transport of spin down electrons is due to evanescent states. As a consequence, the transmission obtained from the DFT+NEGF calculations differs by several order of magnitude for the two spin channels, resulting in the almost perfect spin-polarization of F4GT.

 \section{DOS and orbital occupations}
\begin{figure}[t]
\includegraphics[width=1.0\linewidth]{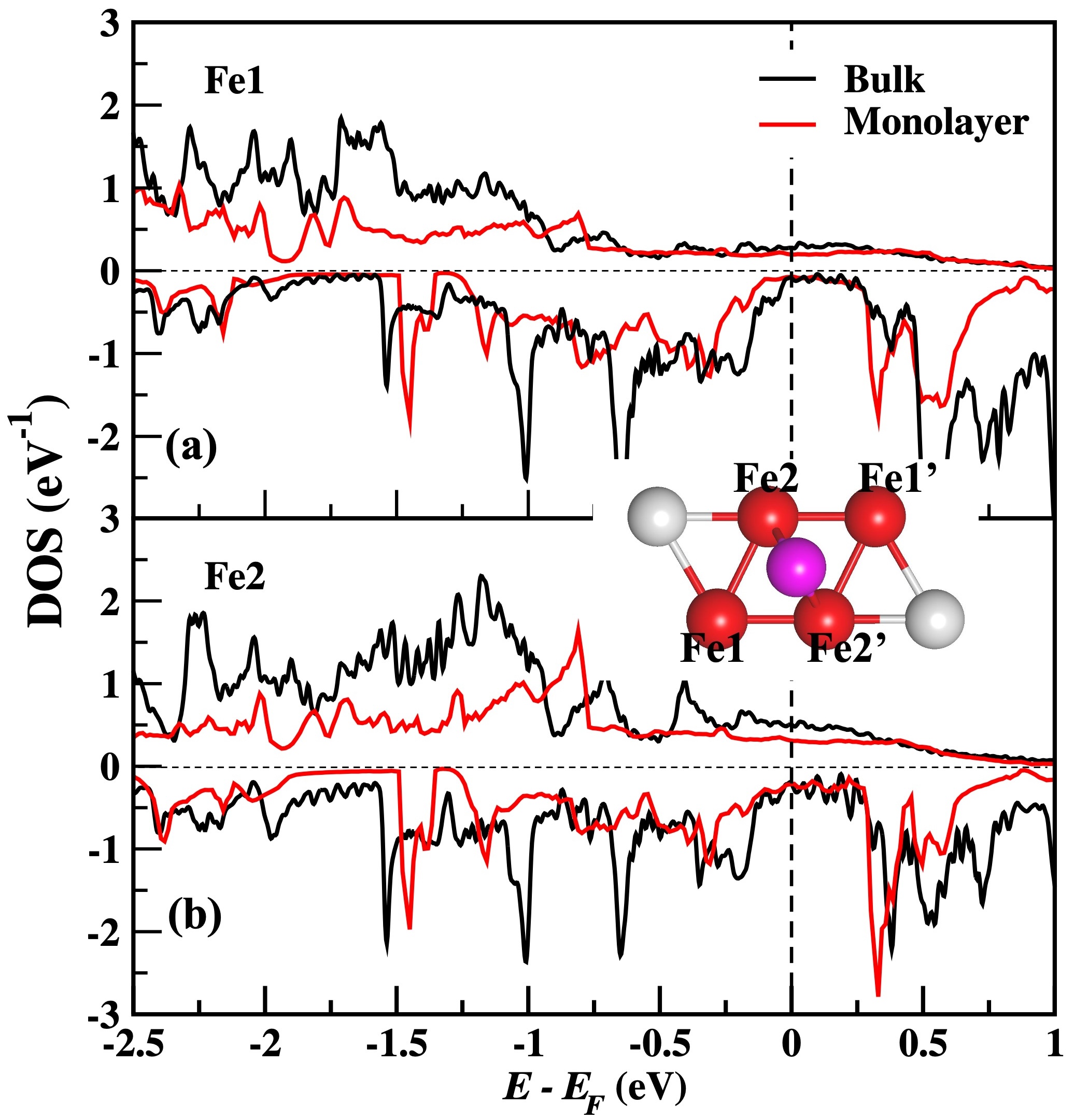}
\caption{Zero-bias DOS of bulk (black) and monolayer (red) F4GT projected over the Fe atoms. The panels (a) and (b) correspond to the two in-equivalent Fe atoms. 
The inset shows the F4GT monolayer where the in-equivalent Fe atoms are indicated. At zero-bias, Fe1 (Fe2) is equivalent to Fe1' (Fe2'). At finite bias, all Fe atoms, Fe1, Fe2, Fe1', and Fe2', become in-equivalent.
The grey and magenta spheres represent the Te and Ge atoms, respectively.}
\label{SI2}
\end{figure}

Fig. \ref{SI2} displays the zero-bias spin-polarized DOS of the bulk and monolayer F4GT device, projected onto the two in-equivalent Fe atoms, Fe1 and Fe2. 

In the bulk case, Fe2 has a stronger ferromagnetic character than Fe1. The spin up $3d$ PDOS at the Fermi level is larger for Fe2 than for Fe1. Furthermore, the two atoms also have a different spin-dependent filling. This is seen from the Mulliken populations in Tab. \ref{MullBulk}. The average spin up (spin down) $3d$ orbital occupations are $0.92$ ($0.39$) and $0.84$ ($0.48$) electrons for Fe1 and Fe2, respectively. As such, the overall charge of Fe1 and Fe2 is the same, while their magnetic moments are about 2.7 $\mu_\mathrm{B}$ and 1.8 $\mu_\mathrm{B}$. 

In the monolayer case, we observe a reduction of the spin-splitting of the PDOS for both Fe atoms compared to the bulk case. Consequently, the magnetic moments from the Mulliken populations, which are reported in Tab. \ref{Mull1}, are also slightly reduced. Nonetheless, we see that the magnetic moment of Fe1 still remains about 0.9 $\mu_\mathrm{B}$ larger than the one of Fe2. This finding is consistent with the results of previous DFT calculations in the literature \cite{FGT4-2022-localmom, bsanyal} (see also Sec. \ref{sec.magn_mom}). 

\begin{table}[h]
    \centering

    \begin{tabular}{|c|c|c|c|c|c|c|}
        \hline
          & $3d_{xy}$  &   $3d_{yz}$  &  $3d_{z^2}$   &  $3d_{xz}$   & $3d_{x^2-y^2}$  & $m$ \\
           & $\uparrow$/$\downarrow$ &   $\uparrow$/$\downarrow$ &   $\uparrow$/$\downarrow$  &  $\uparrow$/$\downarrow$  & $\uparrow$/$\downarrow$ & $ (\mu_\mathrm{B})$ \\     
        \hline
        Fe1 & 0.93/0.33 & 0.88/0.39  & 0.96/0.5 & 0.88/0.39 & 0.93/0.33 & 2.72\\
        Fe2 & 0.85/0.45 & 0.85/0.47 & 0.8/0.54 & 0.85/0.47 &0.85/0.45 & 1.84\\
        \hline
       
    \end{tabular}
     \caption{Magnetic moment, $m$, and spin up/down Mulliken populations of the Fe1 and Fe2 $3d$ orbitals for bulk F4GT at zero bias.}

\label{MullBulk}
\end{table}

\begin{table}[h]
    \centering

    \begin{tabular}{|c|c|c|c|c|c|c|}
        \hline
          & $3d_{xy}$  &   $3d_{yz}$  &  $3d_{z^2}$   &  $3d_{xz}$   & $3d_{x^2-y^2}$  & $m$ \\
           & $\uparrow$/$\downarrow$ &   $\uparrow$/$\downarrow$ &   $\uparrow$/$\downarrow$  &  $\uparrow$/$\downarrow$  & $\uparrow$/$\downarrow$ & $ (\mu_\mathrm{B})$ \\     
        \hline
        Fe1 & 0.93/0.35  &0.87/0.41  & 0.93/0.46 & 0.86/0.41 & 0.93/0.35 & 2.63 \\
        Fe2 &  0.85/0.47  & 0.85/0.48 & 0.81/0.5 & 0.84/0.49  & 0.84/0.47 & 1.75 \\
        \hline
       
    \end{tabular}
     \caption{Magnetic moment, $m$, and spin up/down Mulliken populations of the Fe1 and Fe2 $3d$ orbitals for the F4GT monolayer at zero bias.}

\label{Mull1}
\end{table}


\begin{figure}[t]
\includegraphics[width=1.0\linewidth]{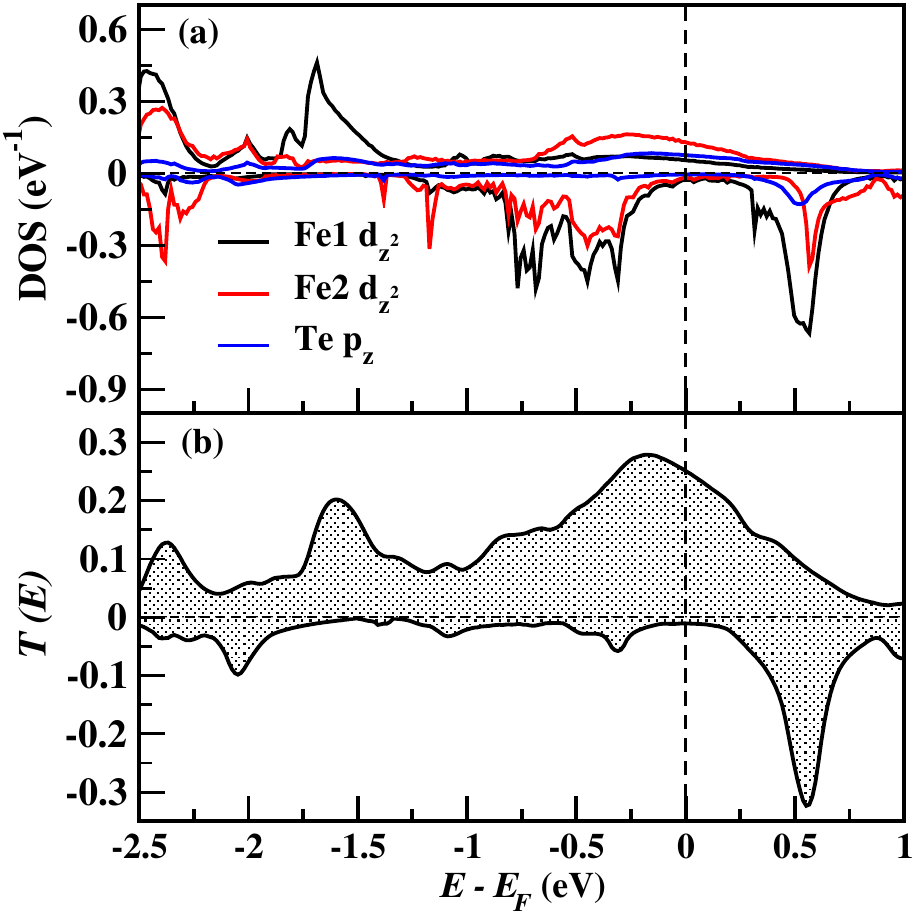}
\caption{(a) Zero bias PDOS projected over $3d_{z^2}$ orbitals of Fe1 (black), Fe2 (red) and Te (blue) $5p_{z}$, (b) Zero bias transmission coefficient for monolayer.}
\label{FigPDOS}
\end{figure}

\section{PDOS analysis for the F4GT monolayer}\label{Sec.PDOS}
Fe1, Fe2 and Te in an F4GT layer are aligned along the $z$ transport direction in a chain-like fashion [see the inset of Fig. \ref{SI2} and Fig. 1(b) in the paper]. Hence, the head-on overlap of their orbitals, forming $\sigma$ covalent bonds, determine the effective delocalization of the electronic states, their coupling to the leads, and ultimately the transport properties. To see that, we present in Fig. \ref{FigPDOS}(a) the zero bias DOS projected over the $3d_{z^2}$ orbital of Fe1 (black curve) and Fe2 (red curve) and the $5p_z$ orbital of the Te atoms (blue curve), which, in the monolayer device, points towards the leads. We can then establish a clear correspondence between the PDOS and the zero-bias transmission coefficient, $T^\sigma(E)$, which is displayed Fig. 1(d) of the paper and which we show again in Fig. \ref{FigPDOS}(b) in the interest of clarity.

For spin up, the Fe $3d_{z^2}$ and the Te $5p_z$ PDOS strongly overlap forming a very broad resonance centered at $E-E_\mathrm{F}\approx -0.25$ eV. This means that the electronic states within that energy region extend across the whole F4GT layer's thickness. Furthermore, owing to their significant contribution from the Te $5p_z$ orbitals, these states can also couple to the leads. This is the most ideal situation for transport as electrons can traverse through the devices with minimal scattering. Accordingly, the zero-bias spin up transmission is found to have a pronounced peak with the same shape as the resonance in the PDOS at $E-E_\mathrm{F}\approx -0.25$ eV. 

The spin down channel is gapped at $E_\mathrm{F}$. The spin down conduction states are located in energy at about $E-E_\mathrm{F}\approx 0.5$ eV, while the valence states span from about $E-E_\mathrm{F}\approx -0.25$ to $-0.8$ eV. Crucially, these valence and conduction states possess a different character.

As indicated by the PDOS, the conduction states have contributions from the Fe $3d_{z^2}$ as well as the Te $5p_z$ orbitals. As such, they are quite delocalized, and can couple to the leads, and are expected to result in large conductance. In fact, we observe that the transmission coefficient displays a sharp peak at $E-E_\mathrm{F}\approx 0.5$ eV, which is as high as the transmission resonance in the spin up channel. 

The valence states are mostly localized over the Fe core of the F4GT layer, as evident from  the large Fe PDOS and the vanishing Te $5p_z$ PDOS between $E-E_\mathrm{F}\approx -0.25$ and $-0.8$ eV. Such localization translates in a low conductance. Accordingly, we see that the peak corresponding to the valence states at $E-E_\mathrm{F}\approx -0.3$ eV in the spin down transmission coefficient is an order of magnitude smaller than the peak corresponding to the conduction states. Thus, there is a marked asymmetry of the transmission coefficient with respect to the center of the gap at $E_F$.  

The asymmetry between the conduction and valence states is also highlighted in the paper. It might be important, because, in practical terms, it implies that the nearly half-metallic transport property of F4GT is maintained under p-doping, whereas it might get suppressed under n-doping. Furthermore, the small conductance of the valence states is also fundamental for preserving the almost perfect spin-filtering properties of F4GT in the presence of correlations, as discussed in the paper when presenting the DFT+DMFT results.


 \section{Electronic structure of the F4GT monolayer at finite bias}

\begin{figure}[h]
\includegraphics[width=1.0\linewidth]{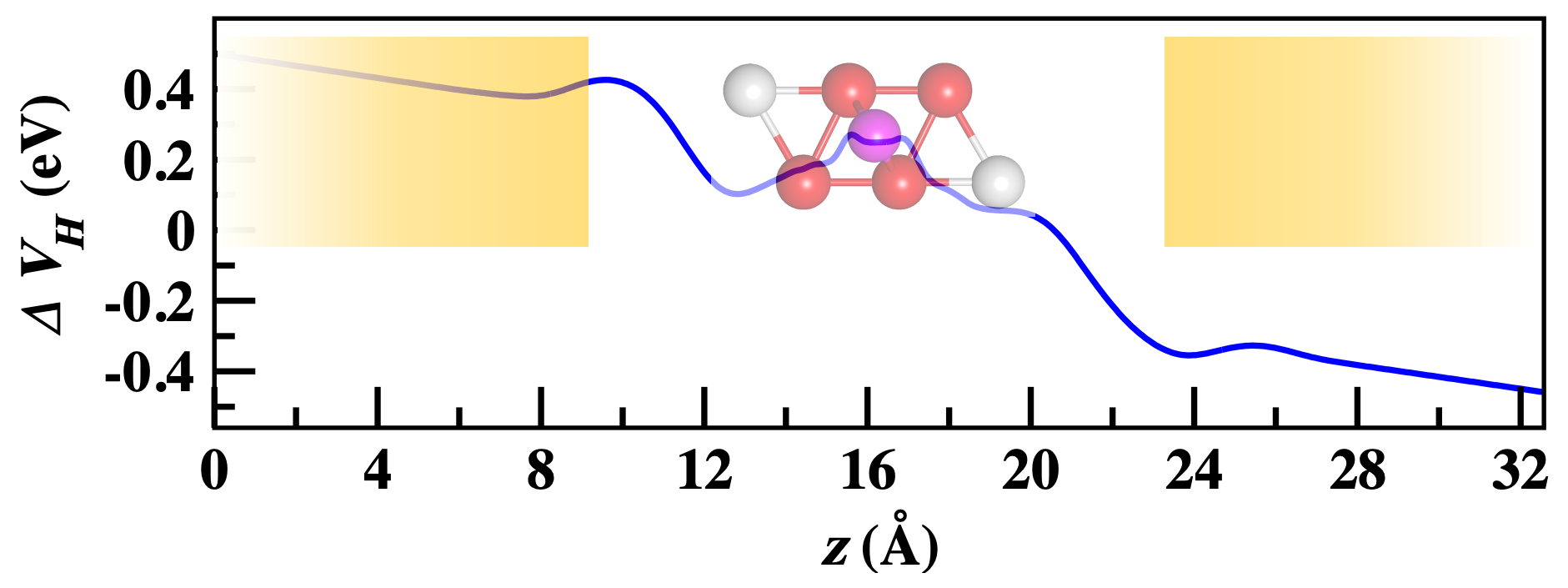}
\caption{Electrostatic potential $\Delta V_\mathrm{H}(z)$ across the central region of the monolayer device under a bias voltage, $V=1$ V. The F4GT layer and the leads are represented in the background as a guide for the eyes. The red, grey, and magenta spheres represent the Fe, Te, and Ge atoms, respectively.}
\label{biasdrop}
\end{figure}

The electronic structure of the F4GT monolayer considerably changes under an applied bias voltage. Specifically, the potential drop, $\Delta V_\mathrm{H}(z)$, across the device's central region, shown in Fig. \ref{biasdrop}, causes some intra-atomic charge redistribution, with all Fe atoms becoming in-equivalent as they are located at a different $z$ coordinate along the device. This effect can be analyzed by comparing the Mulliken populations in Tab. \ref{Mull1} (zero bias) with the ones in Tab. \ref{Mull2} (finite bias, $V=1$ V). The electron occupations of all atoms remain more or less constant. However, a fraction of the electron charge is transferred from the spin up channel to the spin down channel with $V$. Hence, there is a reduction of the Fe atoms' magnetic moments. This effect is particularly evident for Fe2 and Fe2'. At $V=1$ V, their magnetic moments become smaller than the corresponding zero bias values by as much as $\sim 0.7$ $\mu_B$. In practice, the voltage bias modulates the magnetic properties of F4GT. 
\begin{figure}[h]
\includegraphics[width=1.0\linewidth]{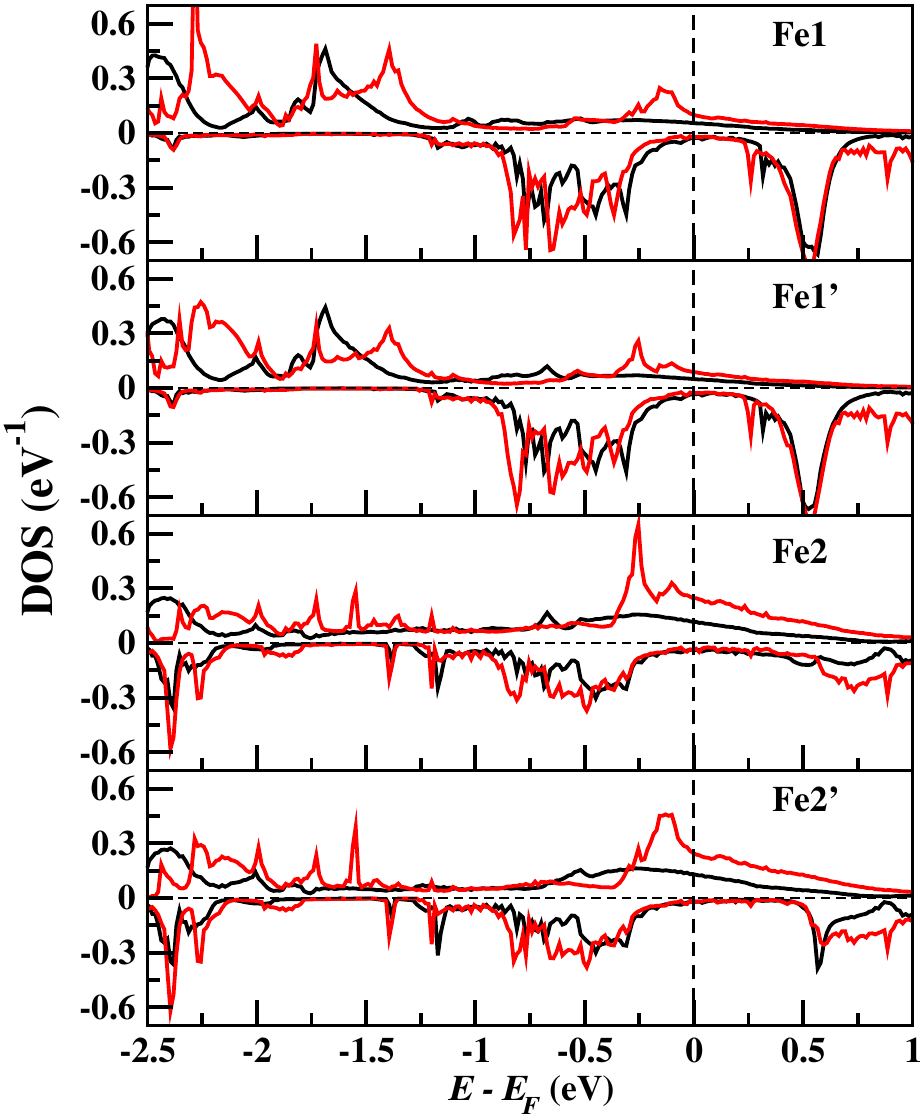}
\caption{$d_{z^2}$-PDOS of the monolayer Fe atoms at zero bias (black) and at $V = 1 $ eV (red).}
\label{SI1}
\end{figure}
We may expect that the reduction of the Fe atoms' magnetic moments would be accompanied by the drastic drop of the spin polarization at $E_\mathrm{F}$. However, that is not actually the case, as we see in the transmission coefficient in Fig. 1 (d) of the paper. To understand this non-trivial finding, we compare the Fe atoms' $d_{z^2}$ PDOS at $V=0$ V and $V=1$ V in Fig. \ref{SI1}. In the spin up channel, the bias induces a narrowing of the central conduction resonance, which is moreover shifted towards the Fermi energy, reflecting the reduction of the spin up electron filling. This effect is compensated in the spin down channel an enhancement as well as a broadening of the valence resonance at $E-E_\mathrm{F}\approx -0.5$, accounting for the increasing spin down occupation. In spite of that, the gap across $E_\mathrm{F}$, and therefore the half-metallic character of the system, remains unchanged.

\begin{table}[h]
    \centering

    \begin{tabular}{|c|c|c|c|c|c|c|}
        \hline
          & $3d_{xy}$  &   $3d_{yz}$  &  $3d_{z^2}$   &  $3_{dxz}$   & $3d_{x^2-y^2}$  & $m$ \\
           & $\uparrow$/$\downarrow$ &   $\uparrow$/$\downarrow$ &   $\uparrow$/$\downarrow$  &  $\uparrow$/$\downarrow$  & $\uparrow$/$\downarrow$ & $ (\mu_\mathrm{B})$ \\     
        \hline
        Fe1 & 0.91/0.37 &  0.86/0.42 &  0.89/0.46 &  0.86/0.43 &  0.91/0.37& 2.48 \\
        Fe1'& 0.91/0.37 & 0.87/0.4 & 0.91/0.45 & 0.87/0.4 & 0.92/0.37 &    2.62 \\
        Fe2 &0.77/0.59 & 0.77/0.54 & 0.73/0.56 & 0.75/0.55 & 0.77/0.59  &  0.95 \\
        Fe2'& 0.78/0.59 & 0.78/0.54 & 0.73/0.54 & 0.77/0.55 & 0.78/0.59 &  1.0\\
        \hline
       
    \end{tabular}
     \caption{Magnetic moment, $m$, and spin up/down Mulliken populations of the $3d$ orbitals of all the Fe atoms in the the F4GT monolayer at $V=1$ V.}

\label{Mull2}
\end{table}

\section{Electronic structure of the F4GT monolayer predicted by DFT+U and DFT+DMFT calculations}


\begin{figure}[t]\label{ldau_trc}
\includegraphics[width=1.0\linewidth]{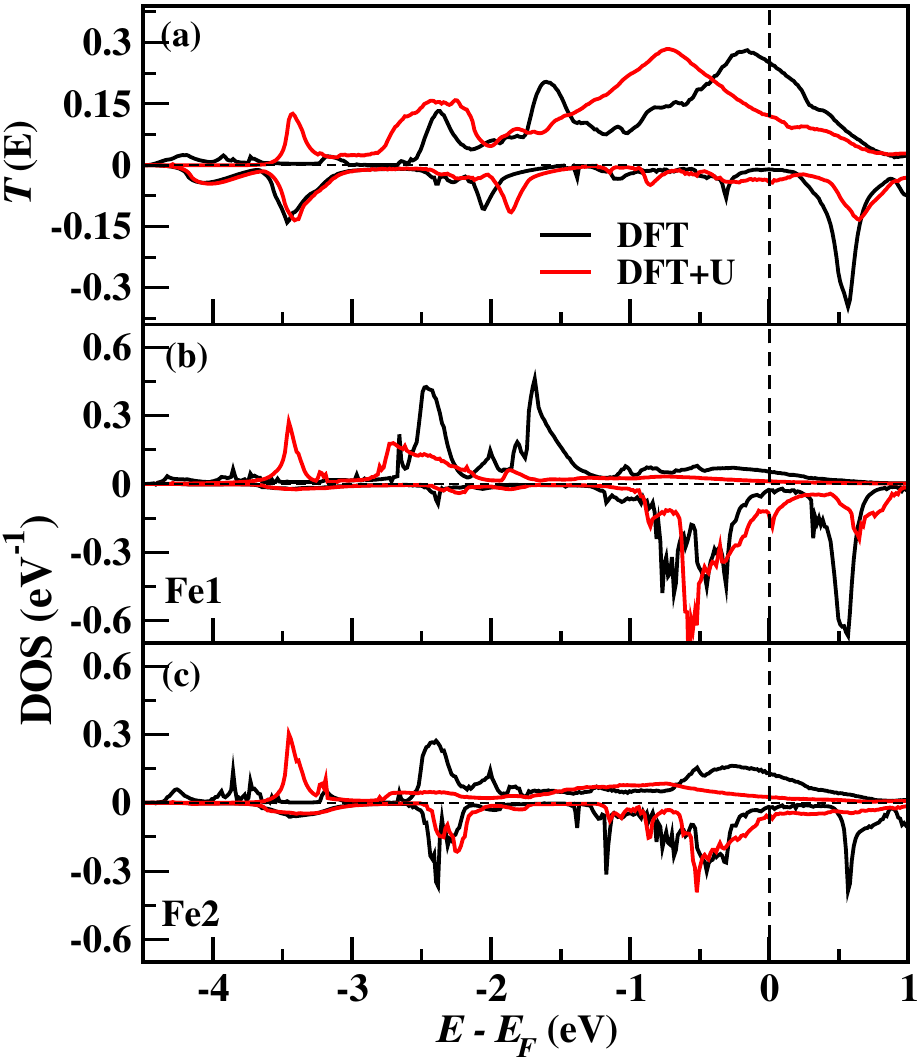}
\caption{DFT versus DFT+U results. (a) Transmission coefficient. (b) Fe1 $d_{z^2}$ PDOS. (c) Fe2 $d_{z^2}$ PDOS. The interaction parameters in DFT+U are $U=3.0$ eV and $J=0.5$ eV for both Fe atoms. }
\label{ldau_trc}
\end{figure}

\subsection{DFT+U results}

 Figs. \ref{ldau_trc}(a), (b), and (c) compare the zero-bias transmission coefficient and the Fe1 and Fe2 $d_{z^2}$ PDOS of the monolayer device calculated by DFT (GGA) and DFT+U. The $U$ and $J$ parameters of DFT+U are respectively equal to $3.0$ eV and $0.5$ eV for all Fe atoms. 

In the spin up channel, the Hubbard $U$-potential in Eq. (\ref{eqn: dudarev}) is negative as the Fe $3d$ orbitals have more than half-filling DFT occupations. Thus, DFT+U shifts the PDOS toward more negative energies, leading to the complete occupation of Fe spin up orbitals for both the in-equivalent Fe atoms. Specifically, we see that the broad resonances in the spin up transmission function are translated in energy from $E_\mathrm{F}$ in DFT to $E-E_\mathrm{F}\sim -0.8$ eV in DFT+U. 

In the spin down channel, the $U$-potential is positive as the $d$ orbitals are less than half-filled. Thus, the $3d$ orbitals move up in energy, further reducing their occupations. In the spin down PDOS, the peaks corresponding to the valence states are seen to cross the Fermi energy, merging with the conduction states, so that the gap closes and the system loses its half-metallic character.  
 

\begin{figure}[th]
\includegraphics[width=1.0\linewidth]{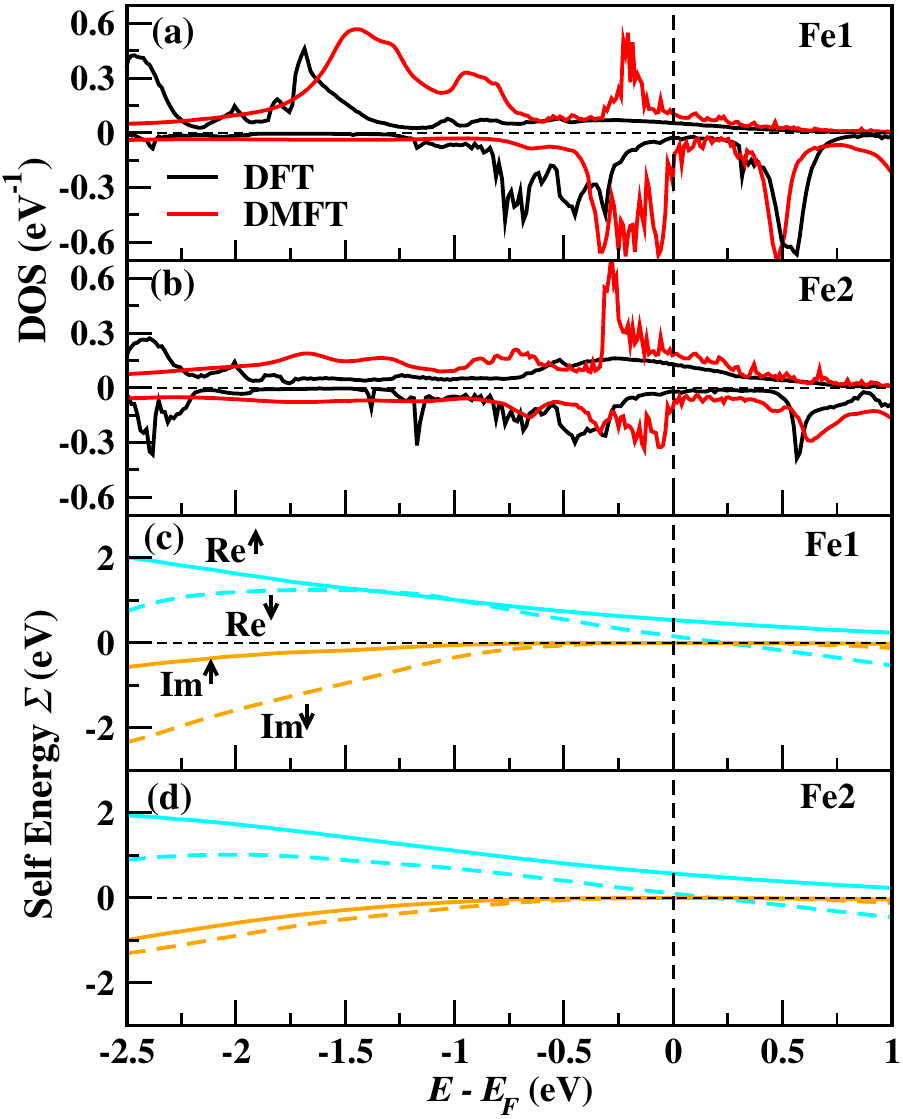}
\caption{$d_{z^2}$ PDOS for Fe1 (a) and Fe2 (b) calculated by DFT (black curve) and DFT+DMFT (red curve). The real (cyan) and imaginary (orange) part of the DMFT self-energy for the $d_{z^2}$ orbitals of Fe1 (c) and Fe2 (d). The interaction parameters in DFT+DMFT are $U=3.0$ eV and $J=0.5$ eV for both Fe atoms.  }
\label{fig: dos_dmft_se}
\end{figure}
\begin{figure}[th]
\includegraphics[width=1.0\linewidth]{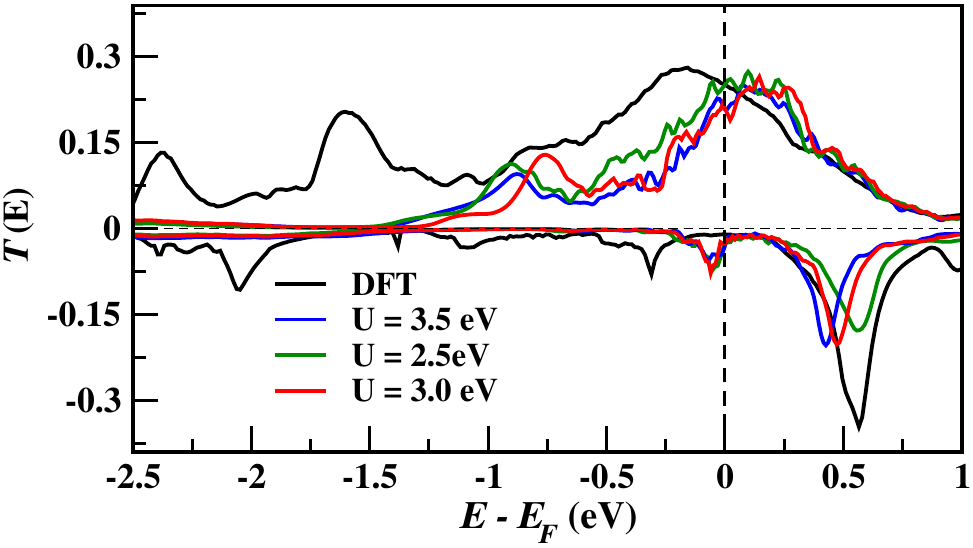}
\caption{Transmission coefficient calculated with DFT and  DFT+DMFT with $U=2.5$ eV, $3.0$ eV, $3.5$ eV, and $J=0.5$ eV. }
\label{fig: dos_dmft_trc}
\end{figure}


\subsection{DFT+DMFT results}
 
The Fe1 and Fe2 $d_{z^2}$ PDOS calculated by DFT (black curve) and DFT+DMFT (red curve) are compared in Figs. \ref{fig: dos_dmft_se}(a) and (b). The real and imaginary parts of the DMFT self-energy of the Fe1 and Fe2 $d_{z^2}$ orbital are plotted in Fig. \ref{fig: dos_dmft_se}(c) and (d). Here, the $U$ and $J$ parameters used for the DFT+DMFT calculations are respectively equal to 3.0 eV and 0.5 eV. The dependence of the results on these parameters is analyzed in Sec. \ref{sec.U}.

In the PDOS plot, we see that DMFT leads to a substantial redistribution of the spectral weight in both spin channels compared to DFT. This effect is due to the real part of the self-energy. Since $\textnormal{Re}[\Sigma^{\sigma}(E)]$ is positive and increases monotonically for $E < E_F$, the states below the Fermi level are drawn toward higher energies. In the spin up channel, the PDOS resonance near the Fermi level gets narrow and sharp. In the spin down channel, the center of valence states resonance is translated up in energy by about $0.3$ eV. Despite that, the resonance's edge remains pinned at the Fermi level without crossing it. The gap in the spin down PDOS is not suppressed, although it is reduced. 

The imaginary part of the DMFT self-energy, $\textnormal{Im}[\Sigma^{\sigma}(E)]$, is typical of a Fermi-liquid, i.e., $\textnormal{Im}[\Sigma^{\sigma}(E)]\propto  (E-E_\mathrm{F})^2$ in both spin channels. Thus, no non-quasi-particle's features are found to emerge inside the gap in the spin down channel. The F4GT monolayer in DFT+DMFT remains half-metallic as in DFT. 


\subsection{Dependence of the DFT+DMFT results on the $U$ interaction parameter}\label{sec.U}

In Fig. \ref{fig: dos_dmft_trc}, we show the zero-bias transmission coefficient of the monolayer device calculated by DFT+DMFT for several values of the local Coulomb interaction, $U =2.5$ eV, $U=3$ eV, and $U=3.5$ eV ($J$ is fixed at 0.5 eV). In all cases, the results look rather similar. In the spin up channel, the main peak of the transmission coefficient is centered around the Fermi level, while, in the spin down channel, there is a marked gap. The size of this gap tends to be reduced with $U$, but the effect is overall quite small. Therefore, the system is found to remain nearly half-metallic in all cases.  



\subsection{Magnetic moments}\label{sec.magn_mom}
Tab. \ref{magmom} reports the magnetic moments calculated by using DFT, DFT+U, and DFT+DMFT for the in-equivalent Fe atoms at zero-bias. The DFT results agree remarkably well with those reported in literature \cite{bsanyal}, but overestimate the experimental values \cite{FGT41st-dup}. DFT+U enhances this overestimation, indicating that the method does not accurately capture electron correlation effects in this system. This trend is consistent with one reported in Ref. [\onlinecite{bsanyal}]. For all values of $U$ considered within DFT+DMFT, we find that the Fe2 moment is enhanced while the Fe1 moment is reduced. This contradicts the findings of Ref. [\onlinecite{bsanyal}], where they observed the opposite trend in their DFT+DMFT calculations: a reduction in the Fe2 moment and enhancement in the Fe1 moment. The origin of this discrepancy may be due to our calculations using the same value of $U$ on both in-equivalent Fe atoms, while Ref. [\onlinecite{bsanyal}] applies an atom specific $U_{\textnormal{Fe1}} > U_{\textnormal{Fe2}}$, obtained from constrained linear-response calculations. This issue is however not particularly important for transport as suggested by the poor dependence of the transmission coefficient on $U$ (see Sec. \ref{sec.U}) and Fig. \ref{fig: dos_dmft_trc}).
We opted for $U=3$ eV for all Fe atoms in our transport calculations due to the excellent agreement between the band structure and the experimental photoemission spectra \cite{antik2}.

\begin{table}
    \centering

    \begin{tabular}{|c|c|c|c|}
        \hline
          & Fe1 & Fe2 & Average \\
        \hline
        DFT & 2.63 & 1.75 & 2.19 \\
        DFT+U ($U=3.0$ eV) & 2.95 & 2.35 & 2.65 \\
        DFT+DMFT($U=2.5$ eV) & 2.58 & 2.02 & 2.30 \\
        DFT+DMFT($U=3.0$ eV) & 2.54 & 2.00 & 2.27 \\
        DFT+DMFT($U=3.5$ eV) & 2.45 & 1.87 & 2.16 \\
        \hline
       
    \end{tabular}
     \caption{Magnetic moments (in $\mu_B$) of the two in-equivalent Fe atoms in the F4GT monolayer at zero bias.}

\label{magmom}
\end{table}

\section{Results of MTJ formed by three F4GT layers}
\begin{figure}[h]
\includegraphics[width=1.0\linewidth]{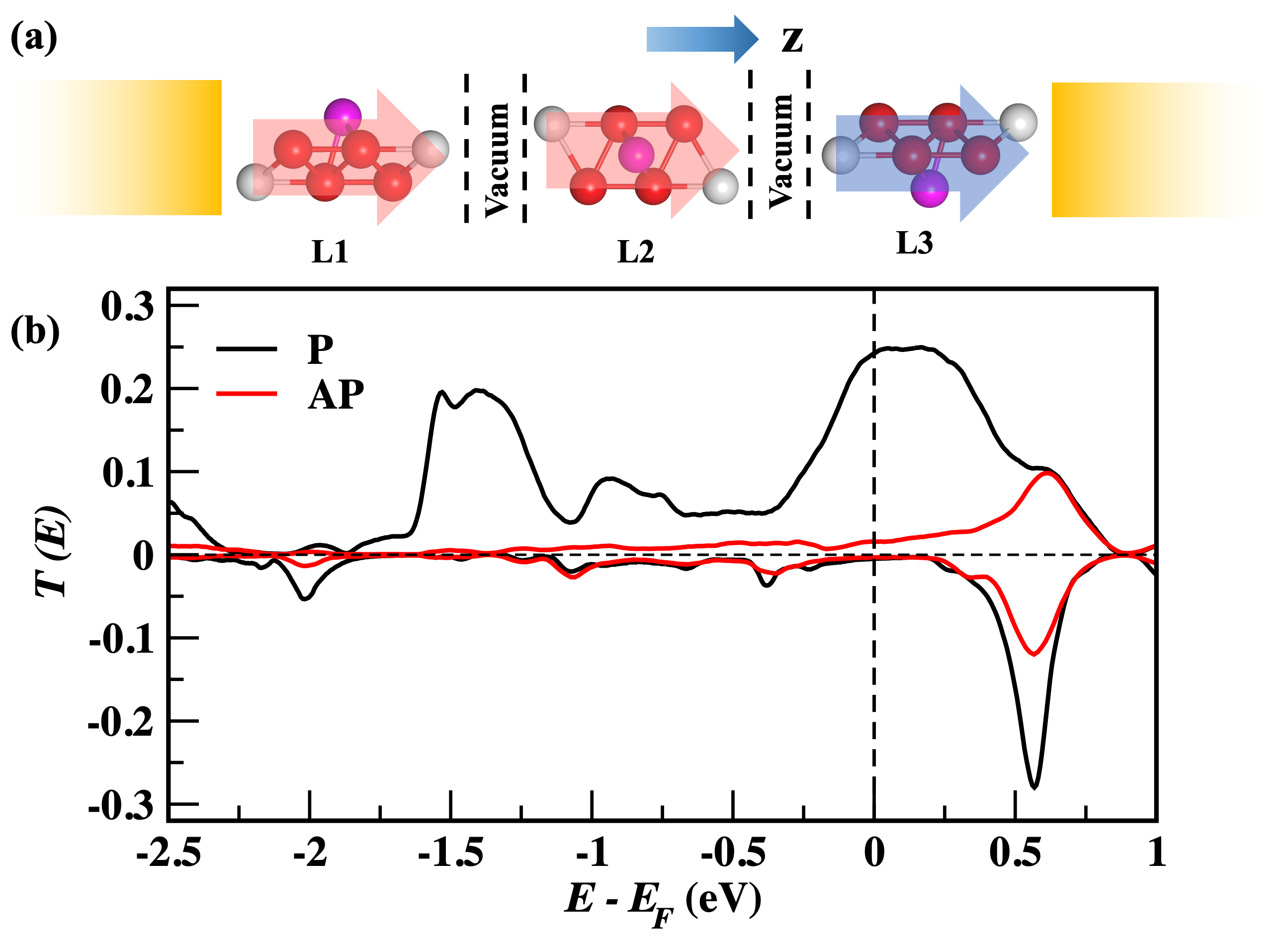}
\caption{Transport calculations for the trilayer MTJ device. (a) Schematic representation of the device. The three F4GT layers are labelled L1, L2 and L3, and are sandwiched between model leads (semi-infinite yellow 
rectangles). (b) Zero bias transmission coefficient for the P and AP configurations.  }
\label{mtj3}
\end{figure}

Here, we present here the results for the device shown in Fig. \ref{mtj3}(a) where the central region consists of three F4GT layers separated by a vdW gap of $3.39$~\AA. The first two layers, denoted as L1 and L2, serve as spin filters, while the third layer, L3, functions as a spin detector. The central region is connected to the same model electrodes as in the monolayer and bilayer devices.

The device can be set in two different magnetic configurations. In the first configuration, the magnetization vectors of three layers are parallel (P) to each other, while in the second, the magnetization vectors of L1 and L2 are antiparallel (AP) to the magnetization vector of L3. The zero-bias P and AP transmission coefficients for both spin channels, $T^\sigma_\textnormal{P}$ and $T^\sigma_\textnormal{AP}$, are shown in Fig. \ref{mtj3}(b). In the P configuration, the transmission coefficient closely resembles that of the monolayer and bilayer devices, exhibiting a half-metallic character. There is a broad resonance around the Fermi level in the spin up channel and a gap in the spin down channel. On the other hand, in the AP configuration, the transmission coefficient is dramatically reduced in both spin channels. Furthermore, $T_\mathrm{AP}^\uparrow(E)$ and $T_\mathrm{AP}^\downarrow(E)$ are not identical. This is because the AP configuration is asymmetric with the magnetization vectors of the first two layers aligned opposite to the magnetization vector of the third layer. In mathematical terms, this asymmetry becomes evident by expressing the spin up and down AP transmission coefficients as square root of the products of the spin up and down transmissions in the P case, namely $T_\mathrm{AP}^\uparrow(E)\approx \sqrt{T^\uparrow_\mathrm{P}(E)T^\uparrow_\mathrm{P}(E)T^\downarrow_\mathrm{P}(E)}$ 
and $T_\mathrm{AP}^\downarrow(E)\approx \sqrt{T^\downarrow_\mathrm{P}(E)T^\downarrow_\mathrm{P}(E)T^\uparrow_\mathrm{P}(E)}$, according to a standard model of MTJs \cite{SanvitoReview,Rungger2020}. 

Finally, the TMR ratio at zero bias, defined as 
\begin{equation}
\mathrm{TMR}=\frac{T_\textnormal{P}(E_\mathrm{F}) - T_{\textnormal{AP}}(E_\mathrm{F})} {T_{\textnormal{AP}}(E_\mathrm{F})}
\end{equation}
with $T_\mathrm{P(AP)}(E)=T^\uparrow_\mathrm{P(AP)}(E)+ T^\downarrow_\mathrm{P(AP)}(E)$, is calculated to reach an impressive $1200 \%$. Such a high predicted value is comparable to the one obtained for Fe(001)/MgO MTJs \cite{bu.zh.01} which is employed in technological applications. Our calculations therefore demonstrate the potential of F4GT for spintronics.

\bibliography{main}

\begin{thebibliography}{80}
\expandafter\ifx\csname natexlab\endcsname\relax\def\natexlab#1{#1}\fi
\expandafter\ifx\csname bibnamefont\endcsname\relax
  \def\bibnamefont#1{#1}\fi
\expandafter\ifx\csname bibfnamefont\endcsname\relax
  \def\bibfnamefont#1{#1}\fi
\expandafter\ifx\csname citenamefont\endcsname\relax
  \def\citenamefont#1{#1}\fi
\expandafter\ifx\csname url\endcsname\relax
  \def\url#1{\texttt{#1}}\fi
\expandafter\ifx\csname urlprefix\endcsname\relax\def\urlprefix{URL }\fi
\providecommand{\bibinfo}[2]{#2}
\providecommand{\eprint}[2][]{\url{#2}}

\bibitem[{\citenamefont{Wolf et~al.}(2001)\citenamefont{Wolf, Awschalom, Buhrman, Daughton, von Molnár, Roukes, Chtchelkanova, and Treger}}]{spintronics}
\bibinfo{author}{\bibfnamefont{S.~A.} \bibnamefont{Wolf}}, \bibinfo{author}{\bibfnamefont{D.~D.} \bibnamefont{Awschalom}}, \bibinfo{author}{\bibfnamefont{R.~A.} \bibnamefont{Buhrman}}, \bibinfo{author}{\bibfnamefont{J.~M.} \bibnamefont{Daughton}}, \bibinfo{author}{\bibfnamefont{S.}~\bibnamefont{von Molnár}}, \bibinfo{author}{\bibfnamefont{M.~L.} \bibnamefont{Roukes}}, \bibinfo{author}{\bibfnamefont{A.~Y.} \bibnamefont{Chtchelkanova}}, \bibnamefont{and} \bibinfo{author}{\bibfnamefont{D.~M.} \bibnamefont{Treger}}, \bibinfo{journal}{Science} \textbf{\bibinfo{volume}{294}}, \bibinfo{pages}{1488} (\bibinfo{year}{2001}), \urlprefix\url{https://www.science.org/doi/abs/10.1126/science.1065389}.

\bibitem[{\citenamefont{Parkin et~al.}(2004)\citenamefont{Parkin, Kaiser, Panchula, Rice, Hughes, Samant, and Yang}}]{parkin}
\bibinfo{author}{\bibfnamefont{S.~S.~P.} \bibnamefont{Parkin}}, \bibinfo{author}{\bibfnamefont{C.}~\bibnamefont{Kaiser}}, \bibinfo{author}{\bibfnamefont{A.}~\bibnamefont{Panchula}}, \bibinfo{author}{\bibfnamefont{P.~M.} \bibnamefont{Rice}}, \bibinfo{author}{\bibfnamefont{B.}~\bibnamefont{Hughes}}, \bibinfo{author}{\bibfnamefont{M.}~\bibnamefont{Samant}}, \bibnamefont{and} \bibinfo{author}{\bibfnamefont{S.-H.} \bibnamefont{Yang}}, \bibinfo{journal}{Nature Materials} \textbf{\bibinfo{volume}{3}}, \bibinfo{pages}{862} (\bibinfo{year}{2004}), ISSN \bibinfo{issn}{1476-4660}, \urlprefix\url{https://doi.org/10.1038/nmat1256}.

\bibitem[{\citenamefont{Julliere}(1975)}]{JULLIERE1975}
\bibinfo{author}{\bibfnamefont{M.}~\bibnamefont{Julliere}}, \bibinfo{journal}{Physics Letters A} \textbf{\bibinfo{volume}{54}}, \bibinfo{pages}{225} (\bibinfo{year}{1975}), ISSN \bibinfo{issn}{0375-9601}, \urlprefix\url{https://www.sciencedirect.com/science/article/pii/0375960175901747}.

\bibitem[{\citenamefont{Moodera et~al.}(1995)\citenamefont{Moodera, Kinder, Wong, and Meservey}}]{MTJ-1}
\bibinfo{author}{\bibfnamefont{J.~S.} \bibnamefont{Moodera}}, \bibinfo{author}{\bibfnamefont{L.~R.} \bibnamefont{Kinder}}, \bibinfo{author}{\bibfnamefont{T.~M.} \bibnamefont{Wong}}, \bibnamefont{and} \bibinfo{author}{\bibfnamefont{R.}~\bibnamefont{Meservey}}, \bibinfo{journal}{Phys. Rev. Lett.} \textbf{\bibinfo{volume}{74}}, \bibinfo{pages}{3273} (\bibinfo{year}{1995}), \urlprefix\url{https://link.aps.org/doi/10.1103/PhysRevLett.74.3273}.

\bibitem[{\citenamefont{Yuasa et~al.}(2004)\citenamefont{Yuasa, Nagahama, Fukushima, Suzuki, and Ando}}]{Yuasa2004}
\bibinfo{author}{\bibfnamefont{S.}~\bibnamefont{Yuasa}}, \bibinfo{author}{\bibfnamefont{T.}~\bibnamefont{Nagahama}}, \bibinfo{author}{\bibfnamefont{A.}~\bibnamefont{Fukushima}}, \bibinfo{author}{\bibfnamefont{Y.}~\bibnamefont{Suzuki}}, \bibnamefont{and} \bibinfo{author}{\bibfnamefont{K.}~\bibnamefont{Ando}}, \bibinfo{journal}{Nature Materials} \textbf{\bibinfo{volume}{3}}, \bibinfo{pages}{868} (\bibinfo{year}{2004}), ISSN \bibinfo{issn}{1476-4660}, \urlprefix\url{https://doi.org/10.1038/nmat1257}.

\bibitem[{\citenamefont{Gibertini et~al.}(2019)\citenamefont{Gibertini, Koperski, Morpurgo, and Novoselov}}]{2dmagnets1}
\bibinfo{author}{\bibfnamefont{M.}~\bibnamefont{Gibertini}}, \bibinfo{author}{\bibfnamefont{M.}~\bibnamefont{Koperski}}, \bibinfo{author}{\bibfnamefont{A.~F.} \bibnamefont{Morpurgo}}, \bibnamefont{and} \bibinfo{author}{\bibfnamefont{K.~S.} \bibnamefont{Novoselov}}, \bibinfo{journal}{Nature Nanotechnology} \textbf{\bibinfo{volume}{14}}, \bibinfo{pages}{408} (\bibinfo{year}{2019}), ISSN \bibinfo{issn}{1748-3395}, \urlprefix\url{https://doi.org/10.1038/s41565-019-0438-6}.

\bibitem[{\citenamefont{Burch et~al.}(2018)\citenamefont{Burch, Mandrus, and Park}}]{2dmagnets2}
\bibinfo{author}{\bibfnamefont{K.~S.} \bibnamefont{Burch}}, \bibinfo{author}{\bibfnamefont{D.}~\bibnamefont{Mandrus}}, \bibnamefont{and} \bibinfo{author}{\bibfnamefont{J.-G.} \bibnamefont{Park}}, \bibinfo{journal}{Nature} \textbf{\bibinfo{volume}{563}}, \bibinfo{pages}{47} (\bibinfo{year}{2018}), ISSN \bibinfo{issn}{1476-4687}, \urlprefix\url{https://doi.org/10.1038/s41586-018-0631-z}.

\bibitem[{\citenamefont{Song et~al.}(2018)\citenamefont{Song, Cai, Tu, Zhang, Huang, Wilson, Seyler, Zhu, Taniguchi, Watanabe et~al.}}]{CrI3-MTJ1}
\bibinfo{author}{\bibfnamefont{T.}~\bibnamefont{Song}}, \bibinfo{author}{\bibfnamefont{X.}~\bibnamefont{Cai}}, \bibinfo{author}{\bibfnamefont{M.~W.-Y.} \bibnamefont{Tu}}, \bibinfo{author}{\bibfnamefont{X.}~\bibnamefont{Zhang}}, \bibinfo{author}{\bibfnamefont{B.}~\bibnamefont{Huang}}, \bibinfo{author}{\bibfnamefont{N.~P.} \bibnamefont{Wilson}}, \bibinfo{author}{\bibfnamefont{K.~L.} \bibnamefont{Seyler}}, \bibinfo{author}{\bibfnamefont{L.}~\bibnamefont{Zhu}}, \bibinfo{author}{\bibfnamefont{T.}~\bibnamefont{Taniguchi}}, \bibinfo{author}{\bibfnamefont{K.}~\bibnamefont{Watanabe}}, \bibnamefont{et~al.}, \bibinfo{journal}{Science} \textbf{\bibinfo{volume}{360}}, \bibinfo{pages}{1214} (\bibinfo{year}{2018}), \eprint{https://www.science.org/doi/pdf/10.1126/science.aar4851}, \urlprefix\url{https://www.science.org/doi/abs/10.1126/science.aar4851}.

\bibitem[{\citenamefont{Klein et~al.}(2018)\citenamefont{Klein, MacNeill, Lado, Soriano, Navarro-Moratalla, Watanabe, Taniguchi, Manni, Canfield, Fernández-Rossier et~al.}}]{CrI3-MTJ2}
\bibinfo{author}{\bibfnamefont{D.~R.} \bibnamefont{Klein}}, \bibinfo{author}{\bibfnamefont{D.}~\bibnamefont{MacNeill}}, \bibinfo{author}{\bibfnamefont{J.~L.} \bibnamefont{Lado}}, \bibinfo{author}{\bibfnamefont{D.}~\bibnamefont{Soriano}}, \bibinfo{author}{\bibfnamefont{E.}~\bibnamefont{Navarro-Moratalla}}, \bibinfo{author}{\bibfnamefont{K.}~\bibnamefont{Watanabe}}, \bibinfo{author}{\bibfnamefont{T.}~\bibnamefont{Taniguchi}}, \bibinfo{author}{\bibfnamefont{S.}~\bibnamefont{Manni}}, \bibinfo{author}{\bibfnamefont{P.}~\bibnamefont{Canfield}}, \bibinfo{author}{\bibfnamefont{J.}~\bibnamefont{Fernández-Rossier}}, \bibnamefont{et~al.}, \bibinfo{journal}{Science} \textbf{\bibinfo{volume}{360}}, \bibinfo{pages}{1218} (\bibinfo{year}{2018}), \eprint{https://www.science.org/doi/pdf/10.1126/science.aar3617}, \urlprefix\url{https://www.science.org/doi/abs/10.1126/science.aar3617}.

\bibitem[{\citenamefont{Seo et~al.}(2020)\citenamefont{Seo, Kim, An, Kim, Kim, Hwang, Kim, Jang, Kim, Eom et~al.}}]{FGT41st-dup}
\bibinfo{author}{\bibfnamefont{J.}~\bibnamefont{Seo}}, \bibinfo{author}{\bibfnamefont{D.~Y.} \bibnamefont{Kim}}, \bibinfo{author}{\bibfnamefont{E.~S.} \bibnamefont{An}}, \bibinfo{author}{\bibfnamefont{K.}~\bibnamefont{Kim}}, \bibinfo{author}{\bibfnamefont{G.-Y.} \bibnamefont{Kim}}, \bibinfo{author}{\bibfnamefont{S.-Y.} \bibnamefont{Hwang}}, \bibinfo{author}{\bibfnamefont{D.~W.} \bibnamefont{Kim}}, \bibinfo{author}{\bibfnamefont{B.~G.} \bibnamefont{Jang}}, \bibinfo{author}{\bibfnamefont{H.}~\bibnamefont{Kim}}, \bibinfo{author}{\bibfnamefont{G.}~\bibnamefont{Eom}}, \bibnamefont{et~al.}, \bibinfo{journal}{Science Advances} \textbf{\bibinfo{volume}{6}}, \bibinfo{pages}{eaay8912} (\bibinfo{year}{2020}), \eprint{https://www.science.org/doi/pdf/10.1126/sciadv.aay8912}, \urlprefix\url{https://www.science.org/doi/abs/10.1126/sciadv.aay8912}.

\bibitem[{\citenamefont{Min et~al.}(2022)\citenamefont{Min, Lee, Choi, Lee, Seo, Kim, Ko, Watanabe, Taniguchi, Ha et~al.}}]{hBN}
\bibinfo{author}{\bibfnamefont{K.-H.} \bibnamefont{Min}}, \bibinfo{author}{\bibfnamefont{D.~H.} \bibnamefont{Lee}}, \bibinfo{author}{\bibfnamefont{S.-J.} \bibnamefont{Choi}}, \bibinfo{author}{\bibfnamefont{I.-H.} \bibnamefont{Lee}}, \bibinfo{author}{\bibfnamefont{J.}~\bibnamefont{Seo}}, \bibinfo{author}{\bibfnamefont{D.~W.} \bibnamefont{Kim}}, \bibinfo{author}{\bibfnamefont{K.-T.} \bibnamefont{Ko}}, \bibinfo{author}{\bibfnamefont{K.}~\bibnamefont{Watanabe}}, \bibinfo{author}{\bibfnamefont{T.}~\bibnamefont{Taniguchi}}, \bibinfo{author}{\bibfnamefont{D.~H.} \bibnamefont{Ha}}, \bibnamefont{et~al.}, \bibinfo{journal}{Nature Materials} \textbf{\bibinfo{volume}{21}}, \bibinfo{pages}{1144} (\bibinfo{year}{2022}), ISSN \bibinfo{issn}{1476-4660}, \urlprefix\url{https://doi.org/10.1038/s41563-022-01320-3}.

\bibitem[{\citenamefont{Wang et~al.}(2018)\citenamefont{Wang, Sapkota, Taniguchi, Watanabe, Mandrus, and Morpurgo}}]{hbn2}
\bibinfo{author}{\bibfnamefont{Z.}~\bibnamefont{Wang}}, \bibinfo{author}{\bibfnamefont{D.}~\bibnamefont{Sapkota}}, \bibinfo{author}{\bibfnamefont{T.}~\bibnamefont{Taniguchi}}, \bibinfo{author}{\bibfnamefont{K.}~\bibnamefont{Watanabe}}, \bibinfo{author}{\bibfnamefont{D.}~\bibnamefont{Mandrus}}, \bibnamefont{and} \bibinfo{author}{\bibfnamefont{A.~F.} \bibnamefont{Morpurgo}}, \bibinfo{journal}{Nano Letters} \textbf{\bibinfo{volume}{18}}, \bibinfo{pages}{4303} (\bibinfo{year}{2018}), \bibinfo{note}{pMID: 29870263}, \urlprefix\url{https://doi.org/10.1021/acs.nanolett.8b01278}.

\bibitem[{\citenamefont{Albarakati et~al.}(2019)\citenamefont{Albarakati, Tan, Chen, Partridge, Zheng, Farrar, Mayes, Field, Lee, Wang et~al.}}]{graphite}
\bibinfo{author}{\bibfnamefont{S.}~\bibnamefont{Albarakati}}, \bibinfo{author}{\bibfnamefont{C.}~\bibnamefont{Tan}}, \bibinfo{author}{\bibfnamefont{Z.-J.} \bibnamefont{Chen}}, \bibinfo{author}{\bibfnamefont{J.~G.} \bibnamefont{Partridge}}, \bibinfo{author}{\bibfnamefont{G.}~\bibnamefont{Zheng}}, \bibinfo{author}{\bibfnamefont{L.}~\bibnamefont{Farrar}}, \bibinfo{author}{\bibfnamefont{E.~L.~H.} \bibnamefont{Mayes}}, \bibinfo{author}{\bibfnamefont{M.~R.} \bibnamefont{Field}}, \bibinfo{author}{\bibfnamefont{C.}~\bibnamefont{Lee}}, \bibinfo{author}{\bibfnamefont{Y.}~\bibnamefont{Wang}}, \bibnamefont{et~al.}, \bibinfo{journal}{Science Advances} \textbf{\bibinfo{volume}{5}}, \bibinfo{pages}{eaaw0409} (\bibinfo{year}{2019}), \eprint{https://www.science.org/doi/pdf/10.1126/sciadv.aaw0409}, \urlprefix\url{https://www.science.org/doi/abs/10.1126/sciadv.aaw0409}.

\bibitem[{\citenamefont{Lin et~al.}(2020)\citenamefont{Lin, Yan, Hu, Lv, Zhu, Wang, Wei, Chang, and Wang}}]{MoS2}
\bibinfo{author}{\bibfnamefont{H.}~\bibnamefont{Lin}}, \bibinfo{author}{\bibfnamefont{F.}~\bibnamefont{Yan}}, \bibinfo{author}{\bibfnamefont{C.}~\bibnamefont{Hu}}, \bibinfo{author}{\bibfnamefont{Q.}~\bibnamefont{Lv}}, \bibinfo{author}{\bibfnamefont{W.}~\bibnamefont{Zhu}}, \bibinfo{author}{\bibfnamefont{Z.}~\bibnamefont{Wang}}, \bibinfo{author}{\bibfnamefont{Z.}~\bibnamefont{Wei}}, \bibinfo{author}{\bibfnamefont{K.}~\bibnamefont{Chang}}, \bibnamefont{and} \bibinfo{author}{\bibfnamefont{K.}~\bibnamefont{Wang}}, \bibinfo{journal}{ACS Applied Materials \& Interfaces} \textbf{\bibinfo{volume}{12}}, \bibinfo{pages}{43921} (\bibinfo{year}{2020}), \bibinfo{note}{pMID: 32878440}, \eprint{https://doi.org/10.1021/acsami.0c12483}, \urlprefix\url{https://doi.org/10.1021/acsami.0c12483}.

\bibitem[{\citenamefont{Zhu et~al.}(2021)\citenamefont{Zhu, Lin, Yan, Hu, Wang, Zhao, Deng, Kudrynskyi, Zhou, Kovalyuk et~al.}}]{InSe}
\bibinfo{author}{\bibfnamefont{W.}~\bibnamefont{Zhu}}, \bibinfo{author}{\bibfnamefont{H.}~\bibnamefont{Lin}}, \bibinfo{author}{\bibfnamefont{F.}~\bibnamefont{Yan}}, \bibinfo{author}{\bibfnamefont{C.}~\bibnamefont{Hu}}, \bibinfo{author}{\bibfnamefont{Z.}~\bibnamefont{Wang}}, \bibinfo{author}{\bibfnamefont{L.}~\bibnamefont{Zhao}}, \bibinfo{author}{\bibfnamefont{Y.}~\bibnamefont{Deng}}, \bibinfo{author}{\bibfnamefont{Z.~R.} \bibnamefont{Kudrynskyi}}, \bibinfo{author}{\bibfnamefont{T.}~\bibnamefont{Zhou}}, \bibinfo{author}{\bibfnamefont{Z.~D.} \bibnamefont{Kovalyuk}}, \bibnamefont{et~al.}, \bibinfo{journal}{Advanced Materials} \textbf{\bibinfo{volume}{33}}, \bibinfo{pages}{2104658} (\bibinfo{year}{2021}), \eprint{https://onlinelibrary.wiley.com/doi/pdf/10.1002/adma.202104658}, \urlprefix\url{https://onlinelibrary.wiley.com/doi/abs/10.1002/adma.202104658}.

\bibitem[{\citenamefont{Zhu et~al.}(2023)\citenamefont{Zhu, Zhu, Zhou, Zhang, Lin, Cui, Yan, Wang, Deng, Yang et~al.}}]{GaSe}
\bibinfo{author}{\bibfnamefont{W.}~\bibnamefont{Zhu}}, \bibinfo{author}{\bibfnamefont{Y.}~\bibnamefont{Zhu}}, \bibinfo{author}{\bibfnamefont{T.}~\bibnamefont{Zhou}}, \bibinfo{author}{\bibfnamefont{X.}~\bibnamefont{Zhang}}, \bibinfo{author}{\bibfnamefont{H.}~\bibnamefont{Lin}}, \bibinfo{author}{\bibfnamefont{Q.}~\bibnamefont{Cui}}, \bibinfo{author}{\bibfnamefont{F.}~\bibnamefont{Yan}}, \bibinfo{author}{\bibfnamefont{Z.}~\bibnamefont{Wang}}, \bibinfo{author}{\bibfnamefont{Y.}~\bibnamefont{Deng}}, \bibinfo{author}{\bibfnamefont{H.}~\bibnamefont{Yang}}, \bibnamefont{et~al.}, \bibinfo{journal}{Nature Communications} \textbf{\bibinfo{volume}{14}}, \bibinfo{pages}{5371} (\bibinfo{year}{2023}), ISSN \bibinfo{issn}{2041-1723}, \urlprefix\url{https://doi.org/10.1038/s41467-023-41077-0}.

\bibitem[{\citenamefont{Zheng et~al.}(2022)\citenamefont{Zheng, Ma, Yan, Lin, Zhu, Ji, Wang, and Wang}}]{WSe2}
\bibinfo{author}{\bibfnamefont{Y.}~\bibnamefont{Zheng}}, \bibinfo{author}{\bibfnamefont{X.}~\bibnamefont{Ma}}, \bibinfo{author}{\bibfnamefont{F.}~\bibnamefont{Yan}}, \bibinfo{author}{\bibfnamefont{H.}~\bibnamefont{Lin}}, \bibinfo{author}{\bibfnamefont{W.}~\bibnamefont{Zhu}}, \bibinfo{author}{\bibfnamefont{Y.}~\bibnamefont{Ji}}, \bibinfo{author}{\bibfnamefont{R.}~\bibnamefont{Wang}}, \bibnamefont{and} \bibinfo{author}{\bibfnamefont{K.}~\bibnamefont{Wang}}, \bibinfo{journal}{npj 2D Materials and Applications} \textbf{\bibinfo{volume}{6}}, \bibinfo{pages}{62} (\bibinfo{year}{2022}), ISSN \bibinfo{issn}{2397-7132}, \urlprefix\url{https://doi.org/10.1038/s41699-022-00339-z}.

\bibitem[{\citenamefont{Jin et~al.}(2023)\citenamefont{Jin, Zhang, Wu, Yang, Zhang, and Chang}}]{WS}
\bibinfo{author}{\bibfnamefont{W.}~\bibnamefont{Jin}}, \bibinfo{author}{\bibfnamefont{G.}~\bibnamefont{Zhang}}, \bibinfo{author}{\bibfnamefont{H.}~\bibnamefont{Wu}}, \bibinfo{author}{\bibfnamefont{L.}~\bibnamefont{Yang}}, \bibinfo{author}{\bibfnamefont{W.}~\bibnamefont{Zhang}}, \bibnamefont{and} \bibinfo{author}{\bibfnamefont{H.}~\bibnamefont{Chang}}, \bibinfo{journal}{ACS Applied Materials \& Interfaces} \textbf{\bibinfo{volume}{15}}, \bibinfo{pages}{36519} (\bibinfo{year}{2023}), \bibinfo{note}{pMID: 37466234}, \eprint{https://doi.org/10.1021/acsami.3c06167}, \urlprefix\url{https://doi.org/10.1021/acsami.3c06167}.

\bibitem[{\citenamefont{Su et~al.}(2021)\citenamefont{Su, Li, Zhu, Zhang, You, and Tsymbal}}]{tsymbol-mtj}
\bibinfo{author}{\bibfnamefont{Y.}~\bibnamefont{Su}}, \bibinfo{author}{\bibfnamefont{X.}~\bibnamefont{Li}}, \bibinfo{author}{\bibfnamefont{M.}~\bibnamefont{Zhu}}, \bibinfo{author}{\bibfnamefont{J.}~\bibnamefont{Zhang}}, \bibinfo{author}{\bibfnamefont{L.}~\bibnamefont{You}}, \bibnamefont{and} \bibinfo{author}{\bibfnamefont{E.~Y.} \bibnamefont{Tsymbal}}, \bibinfo{journal}{Nano Letters} \textbf{\bibinfo{volume}{21}}, \bibinfo{pages}{175} (\bibinfo{year}{2021}), \bibinfo{note}{pMID: 33264014}, \urlprefix\url{https://doi.org/10.1021/acs.nanolett.0c03452}.

\bibitem[{\citenamefont{Li et~al.}(2019)\citenamefont{Li, Lü, Zhang, You, Su, and Tsymbal}}]{Tsymbol_mtjhbn}
\bibinfo{author}{\bibfnamefont{X.}~\bibnamefont{Li}}, \bibinfo{author}{\bibfnamefont{J.-T.} \bibnamefont{Lü}}, \bibinfo{author}{\bibfnamefont{J.}~\bibnamefont{Zhang}}, \bibinfo{author}{\bibfnamefont{L.}~\bibnamefont{You}}, \bibinfo{author}{\bibfnamefont{Y.}~\bibnamefont{Su}}, \bibnamefont{and} \bibinfo{author}{\bibfnamefont{E.~Y.} \bibnamefont{Tsymbal}}, \bibinfo{journal}{Nano Letters} \textbf{\bibinfo{volume}{19}}, \bibinfo{pages}{5133} (\bibinfo{year}{2019}), \bibinfo{note}{pMID: 31276417}, \eprint{https://doi.org/10.1021/acs.nanolett.9b01506}, \urlprefix\url{https://doi.org/10.1021/acs.nanolett.9b01506}.

\bibitem[{\citenamefont{Li et~al.}(2021)\citenamefont{Li, Frauenheim, and He}}]{mol-MTJ}
\bibinfo{author}{\bibfnamefont{D.}~\bibnamefont{Li}}, \bibinfo{author}{\bibfnamefont{T.}~\bibnamefont{Frauenheim}}, \bibnamefont{and} \bibinfo{author}{\bibfnamefont{J.}~\bibnamefont{He}}, \bibinfo{journal}{ACS Applied Materials \& Interfaces} \textbf{\bibinfo{volume}{13}}, \bibinfo{pages}{36098} (\bibinfo{year}{2021}), \bibinfo{note}{pMID: 34308645}, \eprint{https://doi.org/10.1021/acsami.1c10673}, \urlprefix\url{https://doi.org/10.1021/acsami.1c10673}.

\bibitem[{\citenamefont{Deng et~al.}(2018)\citenamefont{Deng, Yu, Song, Zhang, Wang, Sun, Yi, Wu, Wu, Zhu et~al.}}]{FGT3}
\bibinfo{author}{\bibfnamefont{Y.}~\bibnamefont{Deng}}, \bibinfo{author}{\bibfnamefont{Y.}~\bibnamefont{Yu}}, \bibinfo{author}{\bibfnamefont{Y.}~\bibnamefont{Song}}, \bibinfo{author}{\bibfnamefont{J.}~\bibnamefont{Zhang}}, \bibinfo{author}{\bibfnamefont{N.~Z.} \bibnamefont{Wang}}, \bibinfo{author}{\bibfnamefont{Z.}~\bibnamefont{Sun}}, \bibinfo{author}{\bibfnamefont{Y.}~\bibnamefont{Yi}}, \bibinfo{author}{\bibfnamefont{Y.~Z.} \bibnamefont{Wu}}, \bibinfo{author}{\bibfnamefont{S.}~\bibnamefont{Wu}}, \bibinfo{author}{\bibfnamefont{J.}~\bibnamefont{Zhu}}, \bibnamefont{et~al.}, \bibinfo{journal}{Nature} \textbf{\bibinfo{volume}{563}}, \bibinfo{pages}{94} (\bibinfo{year}{2018}), ISSN \bibinfo{issn}{1476-4687}, \urlprefix\url{https://doi.org/10.1038/s41586-018-0626-9}.

\bibitem[{\citenamefont{May et~al.}(2019)\citenamefont{May, Ovchinnikov, Zheng, Hermann, Calder, Huang, Fei, Liu, Xu, and McGuire}}]{FGT5}
\bibinfo{author}{\bibfnamefont{A.~F.} \bibnamefont{May}}, \bibinfo{author}{\bibfnamefont{D.}~\bibnamefont{Ovchinnikov}}, \bibinfo{author}{\bibfnamefont{Q.}~\bibnamefont{Zheng}}, \bibinfo{author}{\bibfnamefont{R.}~\bibnamefont{Hermann}}, \bibinfo{author}{\bibfnamefont{S.}~\bibnamefont{Calder}}, \bibinfo{author}{\bibfnamefont{B.}~\bibnamefont{Huang}}, \bibinfo{author}{\bibfnamefont{Z.}~\bibnamefont{Fei}}, \bibinfo{author}{\bibfnamefont{Y.}~\bibnamefont{Liu}}, \bibinfo{author}{\bibfnamefont{X.}~\bibnamefont{Xu}}, \bibnamefont{and} \bibinfo{author}{\bibfnamefont{M.~A.} \bibnamefont{McGuire}}, \bibinfo{journal}{ACS Nano} \textbf{\bibinfo{volume}{13}}, \bibinfo{pages}{4436} (\bibinfo{year}{2019}), \bibinfo{note}{pMID: 30865426}, \eprint{https://doi.org/10.1021/acsnano.8b09660}, \urlprefix\url{https://doi.org/10.1021/acsnano.8b09660}.

\bibitem[{\citenamefont{Rana et~al.}(2023)\citenamefont{Rana, Bhakar, G., Bera, Saini, Pradhan, Mondal, Kabir, and Sheet}}]{mukulprb}
\bibinfo{author}{\bibfnamefont{D.}~\bibnamefont{Rana}}, \bibinfo{author}{\bibfnamefont{M.}~\bibnamefont{Bhakar}}, \bibinfo{author}{\bibfnamefont{B.}~\bibnamefont{G.}}, \bibinfo{author}{\bibfnamefont{S.}~\bibnamefont{Bera}}, \bibinfo{author}{\bibfnamefont{N.}~\bibnamefont{Saini}}, \bibinfo{author}{\bibfnamefont{S.~K.} \bibnamefont{Pradhan}}, \bibinfo{author}{\bibfnamefont{M.}~\bibnamefont{Mondal}}, \bibinfo{author}{\bibfnamefont{M.}~\bibnamefont{Kabir}}, \bibnamefont{and} \bibinfo{author}{\bibfnamefont{G.}~\bibnamefont{Sheet}}, \bibinfo{journal}{Phys. Rev. B} \textbf{\bibinfo{volume}{107}}, \bibinfo{pages}{224422} (\bibinfo{year}{2023}), \urlprefix\url{https://link.aps.org/doi/10.1103/PhysRevB.107.224422}.

\bibitem[{\citenamefont{Kohn}(1999)}]{Kohn_nobel}
\bibinfo{author}{\bibfnamefont{W.}~\bibnamefont{Kohn}}, \bibinfo{journal}{Rev. Mod. Phys.} \textbf{\bibinfo{volume}{71}}, \bibinfo{pages}{1253} (\bibinfo{year}{1999}), \urlprefix\url{https://link.aps.org/doi/10.1103/RevModPhys.71.1253}.

\bibitem[{\citenamefont{Datta}(1995)}]{Datta}
\bibinfo{author}{\bibfnamefont{S.}~\bibnamefont{Datta}}, \emph{\bibinfo{title}{Electronic Transport in Mesoscopic Systems}} (\bibinfo{publisher}{Cambridge University Press}, \bibinfo{address}{Cambridge, UK}, \bibinfo{year}{1995}).

\bibitem[{\citenamefont{Pickett and Eschrig}(2007)}]{Pickett_2007}
\bibinfo{author}{\bibfnamefont{W.~E.} \bibnamefont{Pickett}} \bibnamefont{and} \bibinfo{author}{\bibfnamefont{H.}~\bibnamefont{Eschrig}}, \bibinfo{journal}{Journal of Physics: Condensed Matter} \textbf{\bibinfo{volume}{19}}, \bibinfo{pages}{315203} (\bibinfo{year}{2007}), \urlprefix\url{https://dx.doi.org/10.1088/0953-8984/19/31/315203}.

\bibitem[{\citenamefont{Mavropoulos et~al.}(2004)\citenamefont{Mavropoulos, Galanakis, Popescu, and Dederichs}}]{Mavropoulos_2004}
\bibinfo{author}{\bibfnamefont{P.}~\bibnamefont{Mavropoulos}}, \bibinfo{author}{\bibfnamefont{I.}~\bibnamefont{Galanakis}}, \bibinfo{author}{\bibfnamefont{V.}~\bibnamefont{Popescu}}, \bibnamefont{and} \bibinfo{author}{\bibfnamefont{P.~H.} \bibnamefont{Dederichs}}, \bibinfo{journal}{Journal of Physics: Condensed Matter} \textbf{\bibinfo{volume}{16}}, \bibinfo{pages}{S5759} (\bibinfo{year}{2004}), \urlprefix\url{https://dx.doi.org/10.1088/0953-8984/16/48/043}.

\bibitem[{\citenamefont{Katsnelson et~al.}(2008)\citenamefont{Katsnelson, Irkhin, Chioncel, Lichtenstein, and de~Groot}}]{quasi-rev}
\bibinfo{author}{\bibfnamefont{M.~I.} \bibnamefont{Katsnelson}}, \bibinfo{author}{\bibfnamefont{V.~Y.} \bibnamefont{Irkhin}}, \bibinfo{author}{\bibfnamefont{L.}~\bibnamefont{Chioncel}}, \bibinfo{author}{\bibfnamefont{A.~I.} \bibnamefont{Lichtenstein}}, \bibnamefont{and} \bibinfo{author}{\bibfnamefont{R.~A.} \bibnamefont{de~Groot}}, \bibinfo{journal}{Rev. Mod. Phys.} \textbf{\bibinfo{volume}{80}}, \bibinfo{pages}{315} (\bibinfo{year}{2008}), \urlprefix\url{https://link.aps.org/doi/10.1103/RevModPhys.80.315}.

\bibitem[{\citenamefont{Rocha et~al.}(2006)\citenamefont{Rocha, Garc\'{\i}a-Su\'arez, Bailey, Lambert, Ferrer, and Sanvito}}]{ro.ga.06}
\bibinfo{author}{\bibfnamefont{A.~R.} \bibnamefont{Rocha}}, \bibinfo{author}{\bibfnamefont{V.~M.} \bibnamefont{Garc\'{\i}a-Su\'arez}}, \bibinfo{author}{\bibfnamefont{S.}~\bibnamefont{Bailey}}, \bibinfo{author}{\bibfnamefont{C.}~\bibnamefont{Lambert}}, \bibinfo{author}{\bibfnamefont{J.}~\bibnamefont{Ferrer}}, \bibnamefont{and} \bibinfo{author}{\bibfnamefont{S.}~\bibnamefont{Sanvito}}, \bibinfo{journal}{Phys. Rev. B} \textbf{\bibinfo{volume}{73}}, \bibinfo{pages}{085414} (\bibinfo{year}{2006}), \urlprefix\url{https://link.aps.org/doi/10.1103/PhysRevB.73.085414}.

\bibitem[{\citenamefont{Rungger and Sanvito}(2008)}]{ivan_self_energies.ss.08}
\bibinfo{author}{\bibfnamefont{I.}~\bibnamefont{Rungger}} \bibnamefont{and} \bibinfo{author}{\bibfnamefont{S.}~\bibnamefont{Sanvito}}, \bibinfo{journal}{Phys. Rev. B} \textbf{\bibinfo{volume}{78}}, \bibinfo{pages}{035407} (\bibinfo{year}{2008}), \urlprefix\url{https://link.aps.org/doi/10.1103/PhysRevB.78.035407}.

\bibitem[{\citenamefont{Rungger et~al.}(2019)\citenamefont{Rungger, Droghetti, and Stamenova}}]{book1}
\bibinfo{author}{\bibfnamefont{I.}~\bibnamefont{Rungger}}, \bibinfo{author}{\bibfnamefont{A.}~\bibnamefont{Droghetti}}, \bibnamefont{and} \bibinfo{author}{\bibfnamefont{M.}~\bibnamefont{Stamenova}}, in \emph{\bibinfo{booktitle}{Handbook of Materials Modeling. Vol. 1 Methods: Theory and Modeling}}, edited by \bibinfo{editor}{\bibfnamefont{S.}~\bibnamefont{Yip}} \bibnamefont{and} \bibinfo{editor}{\bibfnamefont{W.}~\bibnamefont{W.~Andreoni}} (\bibinfo{publisher}{Springer International Publishing}, \bibinfo{year}{2019}).

\bibitem[{\citenamefont{Soler et~al.}(2002)\citenamefont{Soler, Artacho, Gale, Garc{\'{\i}}a, Junquera, Ordej{\'{o}}n, and S{\'{a}}nchez-Portal}}]{siesta}
\bibinfo{author}{\bibfnamefont{J.~M.} \bibnamefont{Soler}}, \bibinfo{author}{\bibfnamefont{E.}~\bibnamefont{Artacho}}, \bibinfo{author}{\bibfnamefont{J.~D.} \bibnamefont{Gale}}, \bibinfo{author}{\bibfnamefont{A.}~\bibnamefont{Garc{\'{\i}}a}}, \bibinfo{author}{\bibfnamefont{J.}~\bibnamefont{Junquera}}, \bibinfo{author}{\bibfnamefont{P.}~\bibnamefont{Ordej{\'{o}}n}}, \bibnamefont{and} \bibinfo{author}{\bibfnamefont{D.}~\bibnamefont{S{\'{a}}nchez-Portal}}, \bibinfo{journal}{Journal of Physics: Condensed Matter} \textbf{\bibinfo{volume}{14}}, \bibinfo{pages}{2745} (\bibinfo{year}{2002}), \urlprefix\url{https://doi.org/10.1088/0953-8984/14/11/302}.

\bibitem[{\citenamefont{Perdew et~al.}(1996)\citenamefont{Perdew, Burke, and Ernzerhof}}]{PBE}
\bibinfo{author}{\bibfnamefont{J.~P.} \bibnamefont{Perdew}}, \bibinfo{author}{\bibfnamefont{K.}~\bibnamefont{Burke}}, \bibnamefont{and} \bibinfo{author}{\bibfnamefont{M.}~\bibnamefont{Ernzerhof}}, \bibinfo{journal}{Phys. Rev. Lett.} \textbf{\bibinfo{volume}{77}}, \bibinfo{pages}{3865} (\bibinfo{year}{1996}), \urlprefix\url{https://link.aps.org/doi/10.1103/PhysRevLett.77.3865}.

\bibitem[{\citenamefont{Mott and Fowler}(1936)}]{Mott}
\bibinfo{author}{\bibfnamefont{N.~F.} \bibnamefont{Mott}} \bibnamefont{and} \bibinfo{author}{\bibfnamefont{R.~H.} \bibnamefont{Fowler}}, \bibinfo{journal}{Proceedings of the Royal Society of London. Series A - Mathematical and Physical Sciences} \textbf{\bibinfo{volume}{153}}, \bibinfo{pages}{699} (\bibinfo{year}{1936}), \urlprefix\url{https://royalsocietypublishing.org/doi/abs/10.1098/rspa.1936.0031}.

\bibitem[{\citenamefont{Fisher and Lee}(1981)}]{PhysRevB.23.6851}
\bibinfo{author}{\bibfnamefont{D.~S.} \bibnamefont{Fisher}} \bibnamefont{and} \bibinfo{author}{\bibfnamefont{P.~A.} \bibnamefont{Lee}}, \bibinfo{journal}{Phys. Rev. B} \textbf{\bibinfo{volume}{23}}, \bibinfo{pages}{6851} (\bibinfo{year}{1981}), \urlprefix\url{https://link.aps.org/doi/10.1103/PhysRevB.23.6851}.

\bibitem[{\citenamefont{Rungger et~al.}(2009)\citenamefont{Rungger, Mryasov, and Sanvito}}]{IvanFeMgO}
\bibinfo{author}{\bibfnamefont{I.}~\bibnamefont{Rungger}}, \bibinfo{author}{\bibfnamefont{O.}~\bibnamefont{Mryasov}}, \bibnamefont{and} \bibinfo{author}{\bibfnamefont{S.}~\bibnamefont{Sanvito}}, \bibinfo{journal}{Phys. Rev. B} \textbf{\bibinfo{volume}{79}}, \bibinfo{pages}{094414} (\bibinfo{year}{2009}), \urlprefix\url{https://link.aps.org/doi/10.1103/PhysRevB.79.094414}.

\bibitem[{\citenamefont{Landauer}(1957)}]{La.57}
\bibinfo{author}{\bibfnamefont{R.}~\bibnamefont{Landauer}}, \bibinfo{journal}{IBM Journal of Research and Development} \textbf{\bibinfo{volume}{1}}, \bibinfo{pages}{223} (\bibinfo{year}{1957}).

\bibitem[{\citenamefont{B\"uttiker}(1986)}]{Bu.86}
\bibinfo{author}{\bibfnamefont{M.}~\bibnamefont{B\"uttiker}}, \bibinfo{journal}{Phys. Rev. Lett.} \textbf{\bibinfo{volume}{57}}, \bibinfo{pages}{1761} (\bibinfo{year}{1986}), \urlprefix\url{https://link.aps.org/doi/10.1103/PhysRevLett.57.1761}.

\bibitem[{\citenamefont{Buttiker}(1988)}]{Bu.88}
\bibinfo{author}{\bibfnamefont{M.}~\bibnamefont{Buttiker}}, \bibinfo{journal}{IBM Journal of Research and Development} \textbf{\bibinfo{volume}{32}}, \bibinfo{pages}{317} (\bibinfo{year}{1988}).

\bibitem[{\citenamefont{Haug and Jauho}(1996)}]{haug1996quantum}
\bibinfo{author}{\bibfnamefont{H.}~\bibnamefont{Haug}} \bibnamefont{and} \bibinfo{author}{\bibfnamefont{A.}~\bibnamefont{Jauho}}, \emph{\bibinfo{title}{Quantum Kinetics in Transport and Optics of Semiconductors}}, Solid-State Sciences Series (\bibinfo{publisher}{Springer Berlin Heidelberg}, \bibinfo{year}{1996}), ISBN \bibinfo{isbn}{9783540616023}.

\bibitem[{\citenamefont{Fernández-Seivane et~al.}(2006)\citenamefont{Fernández-Seivane, Oliveira, Sanvito, and Ferrer}}]{SOC_onsite}
\bibinfo{author}{\bibfnamefont{L.}~\bibnamefont{Fernández-Seivane}}, \bibinfo{author}{\bibfnamefont{M.~A.} \bibnamefont{Oliveira}}, \bibinfo{author}{\bibfnamefont{S.}~\bibnamefont{Sanvito}}, \bibnamefont{and} \bibinfo{author}{\bibfnamefont{J.}~\bibnamefont{Ferrer}}, \bibinfo{journal}{Journal of Physics: Condensed Matter} \textbf{\bibinfo{volume}{18}}, \bibinfo{pages}{7999} (\bibinfo{year}{2006}), \urlprefix\url{https://dx.doi.org/10.1088/0953-8984/18/34/012}.

\bibitem[{\citenamefont{Todorov}(2002)}]{Todorov_2002}
\bibinfo{author}{\bibfnamefont{T.~N.} \bibnamefont{Todorov}}, \bibinfo{journal}{Journal of Physics: Condensed Matter} \textbf{\bibinfo{volume}{14}}, \bibinfo{pages}{3049} (\bibinfo{year}{2002}), \urlprefix\url{https://dx.doi.org/10.1088/0953-8984/14/11/314}.

\bibitem[{\citenamefont{Droghetti et~al.}(2022{\natexlab{a}})\citenamefont{Droghetti, Rungger, Rubio, and Tokatly}}]{PhysRevB.105.024409}
\bibinfo{author}{\bibfnamefont{A.}~\bibnamefont{Droghetti}}, \bibinfo{author}{\bibfnamefont{I.}~\bibnamefont{Rungger}}, \bibinfo{author}{\bibfnamefont{A.}~\bibnamefont{Rubio}}, \bibnamefont{and} \bibinfo{author}{\bibfnamefont{I.~V.} \bibnamefont{Tokatly}}, \bibinfo{journal}{Phys. Rev. B} \textbf{\bibinfo{volume}{105}}, \bibinfo{pages}{024409} (\bibinfo{year}{2022}{\natexlab{a}}), \urlprefix\url{https://link.aps.org/doi/10.1103/PhysRevB.105.024409}.

\bibitem[{\citenamefont{Bajaj et~al.}(2024)\citenamefont{Bajaj, Gupta, Tokatly, Sanvito, and Droghetti}}]{bajaj2024intrinsic}
\bibinfo{author}{\bibfnamefont{A.}~\bibnamefont{Bajaj}}, \bibinfo{author}{\bibfnamefont{R.}~\bibnamefont{Gupta}}, \bibinfo{author}{\bibfnamefont{I.~V.} \bibnamefont{Tokatly}}, \bibinfo{author}{\bibfnamefont{S.}~\bibnamefont{Sanvito}}, \bibnamefont{and} \bibinfo{author}{\bibfnamefont{A.}~\bibnamefont{Droghetti}}, \emph{\bibinfo{title}{Intrinsic spin hall effect in 5d metals calculated from ab-initio transport theory}} (\bibinfo{year}{2024}), \eprint{2401.01658}.

\bibitem[{\citenamefont{Chioncel et~al.}(2015)\citenamefont{Chioncel, Morari, \"Ostlin, Appelt, Droghetti, Radonji\ifmmode~\acute{c}\else \'{c}\fi{}, Rungger, Vitos, Eckern, and Postnikov}}]{liviu_Cu_Co_dmft}
\bibinfo{author}{\bibfnamefont{L.}~\bibnamefont{Chioncel}}, \bibinfo{author}{\bibfnamefont{C.}~\bibnamefont{Morari}}, \bibinfo{author}{\bibfnamefont{A.}~\bibnamefont{\"Ostlin}}, \bibinfo{author}{\bibfnamefont{W.~H.} \bibnamefont{Appelt}}, \bibinfo{author}{\bibfnamefont{A.}~\bibnamefont{Droghetti}}, \bibinfo{author}{\bibfnamefont{M.~M.} \bibnamefont{Radonji\ifmmode~\acute{c}\else \'{c}\fi{}}}, \bibinfo{author}{\bibfnamefont{I.}~\bibnamefont{Rungger}}, \bibinfo{author}{\bibfnamefont{L.}~\bibnamefont{Vitos}}, \bibinfo{author}{\bibfnamefont{U.}~\bibnamefont{Eckern}}, \bibnamefont{and} \bibinfo{author}{\bibfnamefont{A.~V.} \bibnamefont{Postnikov}}, \bibinfo{journal}{Phys. Rev. B} \textbf{\bibinfo{volume}{92}}, \bibinfo{pages}{054431} (\bibinfo{year}{2015}), \urlprefix\url{https://link.aps.org/doi/10.1103/PhysRevB.92.054431}.

\bibitem[{\citenamefont{Droghetti et~al.}(2022{\natexlab{b}})\citenamefont{Droghetti, Radonji\ifmmode~\acute{c}\else \'{c}\fi{}, Chioncel, and Rungger}}]{andrea_Cu_co}
\bibinfo{author}{\bibfnamefont{A.}~\bibnamefont{Droghetti}}, \bibinfo{author}{\bibfnamefont{M.~c. v.~M.} \bibnamefont{Radonji\ifmmode~\acute{c}\else \'{c}\fi{}}}, \bibinfo{author}{\bibfnamefont{L.}~\bibnamefont{Chioncel}}, \bibnamefont{and} \bibinfo{author}{\bibfnamefont{I.}~\bibnamefont{Rungger}}, \bibinfo{journal}{Phys. Rev. B} \textbf{\bibinfo{volume}{106}}, \bibinfo{pages}{075156} (\bibinfo{year}{2022}{\natexlab{b}}), \urlprefix\url{https://link.aps.org/doi/10.1103/PhysRevB.106.075156}.

\bibitem[{\citenamefont{Xu et~al.}(2020)\citenamefont{Xu, Li, Duan, Zhang, Chen, Kang, Liang, Chen, Xia, Xu et~al.}}]{nonstonerFGT}
\bibinfo{author}{\bibfnamefont{X.}~\bibnamefont{Xu}}, \bibinfo{author}{\bibfnamefont{Y.~W.} \bibnamefont{Li}}, \bibinfo{author}{\bibfnamefont{S.~R.} \bibnamefont{Duan}}, \bibinfo{author}{\bibfnamefont{S.~L.} \bibnamefont{Zhang}}, \bibinfo{author}{\bibfnamefont{Y.~J.} \bibnamefont{Chen}}, \bibinfo{author}{\bibfnamefont{L.}~\bibnamefont{Kang}}, \bibinfo{author}{\bibfnamefont{A.~J.} \bibnamefont{Liang}}, \bibinfo{author}{\bibfnamefont{C.}~\bibnamefont{Chen}}, \bibinfo{author}{\bibfnamefont{W.}~\bibnamefont{Xia}}, \bibinfo{author}{\bibfnamefont{Y.}~\bibnamefont{Xu}}, \bibnamefont{et~al.}, \bibinfo{journal}{Phys. Rev. B} \textbf{\bibinfo{volume}{101}}, \bibinfo{pages}{201104} (\bibinfo{year}{2020}), \urlprefix\url{https://link.aps.org/doi/10.1103/PhysRevB.101.201104}.

\bibitem[{\citenamefont{Ghosh et~al.}(2023)\citenamefont{Ghosh, Ershadrad, Borisov, and Sanyal}}]{bsanyal}
\bibinfo{author}{\bibfnamefont{S.}~\bibnamefont{Ghosh}}, \bibinfo{author}{\bibfnamefont{S.}~\bibnamefont{Ershadrad}}, \bibinfo{author}{\bibfnamefont{V.}~\bibnamefont{Borisov}}, \bibnamefont{and} \bibinfo{author}{\bibfnamefont{B.}~\bibnamefont{Sanyal}}, \bibinfo{journal}{npj Computational Materials} \textbf{\bibinfo{volume}{9}}, \bibinfo{pages}{86} (\bibinfo{year}{2023}), \urlprefix\url{https://doi.org/10.1038/s41524-023-01024-5}.

\bibitem[{\citenamefont{Kotliar et~al.}(2006)\citenamefont{Kotliar, Savrasov, Haule, Oudovenko, Parcollet, and Marianetti}}]{dmft1}
\bibinfo{author}{\bibfnamefont{G.}~\bibnamefont{Kotliar}}, \bibinfo{author}{\bibfnamefont{S.~Y.} \bibnamefont{Savrasov}}, \bibinfo{author}{\bibfnamefont{K.}~\bibnamefont{Haule}}, \bibinfo{author}{\bibfnamefont{V.~S.} \bibnamefont{Oudovenko}}, \bibinfo{author}{\bibfnamefont{O.}~\bibnamefont{Parcollet}}, \bibnamefont{and} \bibinfo{author}{\bibfnamefont{C.~A.} \bibnamefont{Marianetti}}, \bibinfo{journal}{Rev. Mod. Phys.} \textbf{\bibinfo{volume}{78}}, \bibinfo{pages}{865} (\bibinfo{year}{2006}), \urlprefix\url{https://link.aps.org/doi/10.1103/RevModPhys.78.865}.

\bibitem[{\citenamefont{Kotliar and Vollhardt}(2004)}]{dmft2}
\bibinfo{author}{\bibfnamefont{G.}~\bibnamefont{Kotliar}} \bibnamefont{and} \bibinfo{author}{\bibfnamefont{D.}~\bibnamefont{Vollhardt}}, \bibinfo{journal}{Physics Today} \textbf{\bibinfo{volume}{57}}, \bibinfo{pages}{53} (\bibinfo{year}{2004}), ISSN \bibinfo{issn}{0031-9228}, \eprint{https://pubs.aip.org/physicstoday/article-pdf/57/3/53/11004005/53\_1\_online.pdf}, \urlprefix\url{https://doi.org/10.1063/1.1712502}.

\bibitem[{\citenamefont{Anisimov et~al.}(1991)\citenamefont{Anisimov, Zaanen, and Andersen}}]{an.za.91}
\bibinfo{author}{\bibfnamefont{V.~I.} \bibnamefont{Anisimov}}, \bibinfo{author}{\bibfnamefont{J.}~\bibnamefont{Zaanen}}, \bibnamefont{and} \bibinfo{author}{\bibfnamefont{O.~K.} \bibnamefont{Andersen}}, \bibinfo{journal}{Phys. Rev. B} \textbf{\bibinfo{volume}{44}}, \bibinfo{pages}{943} (\bibinfo{year}{1991}), \urlprefix\url{https://link.aps.org/doi/10.1103/PhysRevB.44.943}.

\bibitem[{\citenamefont{Liechtenstein et~al.}(1995)\citenamefont{Liechtenstein, Anisimov, and Zaanen}}]{li.an.95}
\bibinfo{author}{\bibfnamefont{A.~I.} \bibnamefont{Liechtenstein}}, \bibinfo{author}{\bibfnamefont{V.~I.} \bibnamefont{Anisimov}}, \bibnamefont{and} \bibinfo{author}{\bibfnamefont{J.}~\bibnamefont{Zaanen}}, \bibinfo{journal}{Phys. Rev. B} \textbf{\bibinfo{volume}{52}}, \bibinfo{pages}{R5467} (\bibinfo{year}{1995}), \urlprefix\url{https://link.aps.org/doi/10.1103/PhysRevB.52.R5467}.

\bibitem[{\citenamefont{Dudarev et~al.}(1998{\natexlab{a}})\citenamefont{Dudarev, Botton, Savrasov, Humphreys, and Sutton}}]{du.bo.98}
\bibinfo{author}{\bibfnamefont{S.~L.} \bibnamefont{Dudarev}}, \bibinfo{author}{\bibfnamefont{G.~A.} \bibnamefont{Botton}}, \bibinfo{author}{\bibfnamefont{S.~Y.} \bibnamefont{Savrasov}}, \bibinfo{author}{\bibfnamefont{C.~J.} \bibnamefont{Humphreys}}, \bibnamefont{and} \bibinfo{author}{\bibfnamefont{A.~P.} \bibnamefont{Sutton}}, \bibinfo{journal}{Phys. Rev. B} \textbf{\bibinfo{volume}{57}}, \bibinfo{pages}{1505} (\bibinfo{year}{1998}{\natexlab{a}}), \urlprefix\url{https://link.aps.org/doi/10.1103/PhysRevB.57.1505}.

\bibitem[{\citenamefont{Droghetti and Rungger}(2017)}]{andrea_ivan_projection}
\bibinfo{author}{\bibfnamefont{A.}~\bibnamefont{Droghetti}} \bibnamefont{and} \bibinfo{author}{\bibfnamefont{I.}~\bibnamefont{Rungger}}, \bibinfo{journal}{Phys. Rev. B} \textbf{\bibinfo{volume}{95}}, \bibinfo{pages}{085131} (\bibinfo{year}{2017}), \urlprefix\url{https://link.aps.org/doi/10.1103/PhysRevB.95.085131}.

\bibitem[{\citenamefont{Droghetti et~al.}(2022{\natexlab{c}})\citenamefont{Droghetti, Radonji\ifmmode~\acute{c}\else \'{c}\fi{}, Halder, Rungger, and Chioncel}}]{sigma2}
\bibinfo{author}{\bibfnamefont{A.}~\bibnamefont{Droghetti}}, \bibinfo{author}{\bibfnamefont{M.~c. v.~M.} \bibnamefont{Radonji\ifmmode~\acute{c}\else \'{c}\fi{}}}, \bibinfo{author}{\bibfnamefont{A.}~\bibnamefont{Halder}}, \bibinfo{author}{\bibfnamefont{I.}~\bibnamefont{Rungger}}, \bibnamefont{and} \bibinfo{author}{\bibfnamefont{L.}~\bibnamefont{Chioncel}}, \bibinfo{journal}{Phys. Rev. B} \textbf{\bibinfo{volume}{105}}, \bibinfo{pages}{115129} (\bibinfo{year}{2022}{\natexlab{c}}), \urlprefix\url{https://link.aps.org/doi/10.1103/PhysRevB.105.115129}.

\bibitem[{\citenamefont{Sanvito and Rocha}(2006)}]{SanvitoReview}
\bibinfo{author}{\bibfnamefont{S.}~\bibnamefont{Sanvito}} \bibnamefont{and} \bibinfo{author}{\bibfnamefont{A.~R.} \bibnamefont{Rocha}}, \bibinfo{journal}{Journal of Computational and Theoretical Nanoscience} \textbf{\bibinfo{volume}{3}}, \bibinfo{pages}{624} (\bibinfo{year}{2006}), \urlprefix\url{https://doi.org/10.1166/jctn.2006.3047}.

\bibitem[{\citenamefont{Rungger et~al.}(2020)\citenamefont{Rungger, Droghetti, and Stamenova}}]{Rungger2020}
\bibinfo{author}{\bibfnamefont{I.}~\bibnamefont{Rungger}}, \bibinfo{author}{\bibfnamefont{A.}~\bibnamefont{Droghetti}}, \bibnamefont{and} \bibinfo{author}{\bibfnamefont{M.}~\bibnamefont{Stamenova}}, \emph{\bibinfo{title}{Non-equilibrium Green's Function Methods for Spin Transport and Dynamics}} (\bibinfo{publisher}{Springer International Publishing}, \bibinfo{address}{Cham}, \bibinfo{year}{2020}), pp. \bibinfo{pages}{957--983}, ISBN \bibinfo{isbn}{978-3-319-44677-6}, \urlprefix\url{https://doi.org/10.1007/978-3-319-44677-6_75}.

\bibitem[{\citenamefont{Shao et~al.}(2023)\citenamefont{Shao, Jiang, Ding, Zhang, Wang, Xiao, Gurung, Lu, Sun, and Tsymbal}}]{tsymbolprl}
\bibinfo{author}{\bibfnamefont{D.-F.} \bibnamefont{Shao}}, \bibinfo{author}{\bibfnamefont{Y.-Y.} \bibnamefont{Jiang}}, \bibinfo{author}{\bibfnamefont{J.}~\bibnamefont{Ding}}, \bibinfo{author}{\bibfnamefont{S.-H.} \bibnamefont{Zhang}}, \bibinfo{author}{\bibfnamefont{Z.-A.} \bibnamefont{Wang}}, \bibinfo{author}{\bibfnamefont{R.-C.} \bibnamefont{Xiao}}, \bibinfo{author}{\bibfnamefont{G.}~\bibnamefont{Gurung}}, \bibinfo{author}{\bibfnamefont{W.~J.} \bibnamefont{Lu}}, \bibinfo{author}{\bibfnamefont{Y.~P.} \bibnamefont{Sun}}, \bibnamefont{and} \bibinfo{author}{\bibfnamefont{E.~Y.} \bibnamefont{Tsymbal}}, \bibinfo{journal}{Phys. Rev. Lett.} \textbf{\bibinfo{volume}{130}}, \bibinfo{pages}{216702} (\bibinfo{year}{2023}), \urlprefix\url{https://link.aps.org/doi/10.1103/PhysRevLett.130.216702}.

\bibitem[{\citenamefont{Butler et~al.}(2001)\citenamefont{Butler, Zhang, Schulthess, and MacLaren}}]{bu.zh.01}
\bibinfo{author}{\bibfnamefont{W.~H.} \bibnamefont{Butler}}, \bibinfo{author}{\bibfnamefont{X.-G.} \bibnamefont{Zhang}}, \bibinfo{author}{\bibfnamefont{T.~C.} \bibnamefont{Schulthess}}, \bibnamefont{and} \bibinfo{author}{\bibfnamefont{J.~M.} \bibnamefont{MacLaren}}, \bibinfo{journal}{Phys. Rev. B} \textbf{\bibinfo{volume}{63}}, \bibinfo{pages}{054416} (\bibinfo{year}{2001}), \urlprefix\url{https://link.aps.org/doi/10.1103/PhysRevB.63.054416}.

\bibitem[{\citenamefont{Wang et~al.}(2023)\citenamefont{Wang, Lu, Guo, Li, Wu, Li, Xie, Sun, Li, Damas et~al.}}]{off-stoichio}
\bibinfo{author}{\bibfnamefont{H.}~\bibnamefont{Wang}}, \bibinfo{author}{\bibfnamefont{H.}~\bibnamefont{Lu}}, \bibinfo{author}{\bibfnamefont{Z.}~\bibnamefont{Guo}}, \bibinfo{author}{\bibfnamefont{A.}~\bibnamefont{Li}}, \bibinfo{author}{\bibfnamefont{P.}~\bibnamefont{Wu}}, \bibinfo{author}{\bibfnamefont{J.}~\bibnamefont{Li}}, \bibinfo{author}{\bibfnamefont{W.}~\bibnamefont{Xie}}, \bibinfo{author}{\bibfnamefont{Z.}~\bibnamefont{Sun}}, \bibinfo{author}{\bibfnamefont{P.}~\bibnamefont{Li}}, \bibinfo{author}{\bibfnamefont{H.}~\bibnamefont{Damas}}, \bibnamefont{et~al.}, \bibinfo{journal}{Nature Communications} \textbf{\bibinfo{volume}{14}}, \bibinfo{pages}{2483} (\bibinfo{year}{2023}), \urlprefix\url{https://doi.org/10.1038/s41467-023-37917-8}.

\bibitem[{\citenamefont{Yang et~al.}(2021)\citenamefont{Yang, Zhou, Feng, and Yao}}]{FGT4-monolayer}
\bibinfo{author}{\bibfnamefont{X.}~\bibnamefont{Yang}}, \bibinfo{author}{\bibfnamefont{X.}~\bibnamefont{Zhou}}, \bibinfo{author}{\bibfnamefont{W.}~\bibnamefont{Feng}}, \bibnamefont{and} \bibinfo{author}{\bibfnamefont{Y.}~\bibnamefont{Yao}}, \bibinfo{journal}{Phys. Rev. B} \textbf{\bibinfo{volume}{104}}, \bibinfo{pages}{104427} (\bibinfo{year}{2021}), \urlprefix\url{https://link.aps.org/doi/10.1103/PhysRevB.104.104427}.

\bibitem[{\citenamefont{Kresse and Hafner}(1993)}]{VASP}
\bibinfo{author}{\bibfnamefont{G.}~\bibnamefont{Kresse}} \bibnamefont{and} \bibinfo{author}{\bibfnamefont{J.}~\bibnamefont{Hafner}}, \bibinfo{journal}{Phys. Rev. B} \textbf{\bibinfo{volume}{47}}, \bibinfo{pages}{558} (\bibinfo{year}{1993}), \urlprefix\url{https://link.aps.org/doi/10.1103/PhysRevB.47.558}.

\bibitem[{\citenamefont{Grimme et~al.}(2010)\citenamefont{Grimme, Antony, Ehrlich, and Krieg}}]{DFTD3}
\bibinfo{author}{\bibfnamefont{S.}~\bibnamefont{Grimme}}, \bibinfo{author}{\bibfnamefont{J.}~\bibnamefont{Antony}}, \bibinfo{author}{\bibfnamefont{S.}~\bibnamefont{Ehrlich}}, \bibnamefont{and} \bibinfo{author}{\bibfnamefont{H.}~\bibnamefont{Krieg}}, \bibinfo{journal}{The Journal of Chemical Physics} \textbf{\bibinfo{volume}{132}}, \bibinfo{pages}{154104} (\bibinfo{year}{2010}), ISSN \bibinfo{issn}{0021-9606}, \eprint{https://pubs.aip.org/aip/jcp/article-pdf/doi/10.1063/1.3382344/15684000/154104\_1\_online.pdf}, \urlprefix\url{https://doi.org/10.1063/1.3382344}.

\bibitem[{\citenamefont{García et~al.}(2020)\citenamefont{García, Papior, Akhtar, Artacho, Blum, Bosoni, Brandimarte, Brandbyge, Cerdá, Corsetti et~al.}}]{siestapsml}
\bibinfo{author}{\bibfnamefont{A.}~\bibnamefont{García}}, \bibinfo{author}{\bibfnamefont{N.}~\bibnamefont{Papior}}, \bibinfo{author}{\bibfnamefont{A.}~\bibnamefont{Akhtar}}, \bibinfo{author}{\bibfnamefont{E.}~\bibnamefont{Artacho}}, \bibinfo{author}{\bibfnamefont{V.}~\bibnamefont{Blum}}, \bibinfo{author}{\bibfnamefont{E.}~\bibnamefont{Bosoni}}, \bibinfo{author}{\bibfnamefont{P.}~\bibnamefont{Brandimarte}}, \bibinfo{author}{\bibfnamefont{M.}~\bibnamefont{Brandbyge}}, \bibinfo{author}{\bibfnamefont{J.~I.} \bibnamefont{Cerdá}}, \bibinfo{author}{\bibfnamefont{F.}~\bibnamefont{Corsetti}}, \bibnamefont{et~al.}, \bibinfo{journal}{The Journal of Chemical Physics} \textbf{\bibinfo{volume}{152}}, \bibinfo{pages}{204108} (\bibinfo{year}{2020}), ISSN \bibinfo{issn}{0021-9606}, \urlprefix\url{https://doi.org/10.1063/5.0005077}.

\bibitem[{\citenamefont{Troullier and Martins}(1991{\natexlab{a}})}]{pseudopotential_1}
\bibinfo{author}{\bibfnamefont{N.}~\bibnamefont{Troullier}} \bibnamefont{and} \bibinfo{author}{\bibfnamefont{J.~L.} \bibnamefont{Martins}}, \bibinfo{journal}{Phys. Rev. B} \textbf{\bibinfo{volume}{43}}, \bibinfo{pages}{1993} (\bibinfo{year}{1991}{\natexlab{a}}), \urlprefix\url{https://link.aps.org/doi/10.1103/PhysRevB.43.1993}.

\bibitem[{\citenamefont{Troullier and Martins}(1991{\natexlab{b}})}]{pseudopotential_2}
\bibinfo{author}{\bibfnamefont{N.}~\bibnamefont{Troullier}} \bibnamefont{and} \bibinfo{author}{\bibfnamefont{J.~L.} \bibnamefont{Martins}}, \bibinfo{journal}{Phys. Rev. B} \textbf{\bibinfo{volume}{43}}, \bibinfo{pages}{8861} (\bibinfo{year}{1991}{\natexlab{b}}), \urlprefix\url{https://link.aps.org/doi/10.1103/PhysRevB.43.8861}.

\bibitem[{\citenamefont{Artacho et~al.}(1999)\citenamefont{Artacho, Sánchez-Portal, Ordejón, García, and Soler}}]{artacho1999linear}
\bibinfo{author}{\bibfnamefont{E.}~\bibnamefont{Artacho}}, \bibinfo{author}{\bibfnamefont{D.}~\bibnamefont{Sánchez-Portal}}, \bibinfo{author}{\bibfnamefont{P.}~\bibnamefont{Ordejón}}, \bibinfo{author}{\bibfnamefont{A.}~\bibnamefont{García}}, \bibnamefont{and} \bibinfo{author}{\bibfnamefont{J.~M.} \bibnamefont{Soler}}, \bibinfo{journal}{physica status solidi (b)} \textbf{\bibinfo{volume}{215}}, \bibinfo{pages}{809} (\bibinfo{year}{1999}), \urlprefix\url{https://onlinelibrary.wiley.com/doi/abs/10.1002/%28SICI%291521-3951%28199909%29215%3A1%3C809%3A%3AAID-PSSB809%3E3.0.CO%3B2-0}.

\bibitem[{\citenamefont{Junquera et~al.}(2001)\citenamefont{Junquera, Paz, S{\'a}nchez-Portal, and Artacho}}]{junquera2001numerical}
\bibinfo{author}{\bibfnamefont{J.}~\bibnamefont{Junquera}}, \bibinfo{author}{\bibfnamefont{{\'O}.}~\bibnamefont{Paz}}, \bibinfo{author}{\bibfnamefont{D.}~\bibnamefont{S{\'a}nchez-Portal}}, \bibnamefont{and} \bibinfo{author}{\bibfnamefont{E.}~\bibnamefont{Artacho}}, \bibinfo{journal}{Physical Review B} \textbf{\bibinfo{volume}{64}}, \bibinfo{pages}{235111} (\bibinfo{year}{2001}).

\bibitem[{\citenamefont{Rivero et~al.}(2015)\citenamefont{Rivero, Garc{\'\i}a-Su{\'a}rez, Pere{\~n}iguez, Utt, Yang, Bellaiche, Park, Ferrer, and Barraza-Lopez}}]{rivero2015systematic}
\bibinfo{author}{\bibfnamefont{P.}~\bibnamefont{Rivero}}, \bibinfo{author}{\bibfnamefont{V.~M.} \bibnamefont{Garc{\'\i}a-Su{\'a}rez}}, \bibinfo{author}{\bibfnamefont{D.}~\bibnamefont{Pere{\~n}iguez}}, \bibinfo{author}{\bibfnamefont{K.}~\bibnamefont{Utt}}, \bibinfo{author}{\bibfnamefont{Y.}~\bibnamefont{Yang}}, \bibinfo{author}{\bibfnamefont{L.}~\bibnamefont{Bellaiche}}, \bibinfo{author}{\bibfnamefont{K.}~\bibnamefont{Park}}, \bibinfo{author}{\bibfnamefont{J.}~\bibnamefont{Ferrer}}, \bibnamefont{and} \bibinfo{author}{\bibfnamefont{S.}~\bibnamefont{Barraza-Lopez}}, \bibinfo{journal}{Computational Materials Science} \textbf{\bibinfo{volume}{98}}, \bibinfo{pages}{372} (\bibinfo{year}{2015}), ISSN \bibinfo{issn}{0927-0256}, \urlprefix\url{https://www.sciencedirect.com/science/article/pii/S0927025614007940}.

\bibitem[{\citenamefont{Imada et~al.}(1998)\citenamefont{Imada, Fujimori, and Tokura}}]{im.fu.98}
\bibinfo{author}{\bibfnamefont{M.}~\bibnamefont{Imada}}, \bibinfo{author}{\bibfnamefont{A.}~\bibnamefont{Fujimori}}, \bibnamefont{and} \bibinfo{author}{\bibfnamefont{Y.}~\bibnamefont{Tokura}}, \bibinfo{journal}{Rev. Mod. Phys.} \textbf{\bibinfo{volume}{70}}, \bibinfo{pages}{1039} (\bibinfo{year}{1998}).

\bibitem[{\citenamefont{Anisimov and Gunnarsson}(1991)}]{an.gu.91}
\bibinfo{author}{\bibfnamefont{V.~I.} \bibnamefont{Anisimov}} \bibnamefont{and} \bibinfo{author}{\bibfnamefont{O.}~\bibnamefont{Gunnarsson}}, \bibinfo{journal}{Phys. Rev. B} \textbf{\bibinfo{volume}{43}}, \bibinfo{pages}{7570} (\bibinfo{year}{1991}), \urlprefix\url{https://link.aps.org/doi/10.1103/PhysRevB.43.7570}.

\bibitem[{\citenamefont{Dudarev et~al.}(1998{\natexlab{b}})\citenamefont{Dudarev, Botton, Savrasov, Humphreys, and Sutton}}]{dudarev}
\bibinfo{author}{\bibfnamefont{S.~L.} \bibnamefont{Dudarev}}, \bibinfo{author}{\bibfnamefont{G.~A.} \bibnamefont{Botton}}, \bibinfo{author}{\bibfnamefont{S.~Y.} \bibnamefont{Savrasov}}, \bibinfo{author}{\bibfnamefont{C.~J.} \bibnamefont{Humphreys}}, \bibnamefont{and} \bibinfo{author}{\bibfnamefont{A.~P.} \bibnamefont{Sutton}}, \bibinfo{journal}{Phys. Rev. B} \textbf{\bibinfo{volume}{57}}, \bibinfo{pages}{1505} (\bibinfo{year}{1998}{\natexlab{b}}), \urlprefix\url{https://link.aps.org/doi/10.1103/PhysRevB.57.1505}.

\bibitem[{\citenamefont{Kohn}(1959)}]{PhysRev.115.809}
\bibinfo{author}{\bibfnamefont{W.}~\bibnamefont{Kohn}}, \bibinfo{journal}{Phys. Rev.} \textbf{\bibinfo{volume}{115}}, \bibinfo{pages}{809} (\bibinfo{year}{1959}), \urlprefix\url{https://link.aps.org/doi/10.1103/PhysRev.115.809}.

\bibitem[{\citenamefont{Heine}(1963)}]{Heine_1963}
\bibinfo{author}{\bibfnamefont{V.}~\bibnamefont{Heine}}, \bibinfo{journal}{Proceedings of the Physical Society} \textbf{\bibinfo{volume}{81}}, \bibinfo{pages}{300} (\bibinfo{year}{1963}), \urlprefix\url{https://dx.doi.org/10.1088/0370-1328/81/2/311}.

\bibitem[{\citenamefont{Prodan}(2006)}]{PhysRevB.73.035128}
\bibinfo{author}{\bibfnamefont{E.}~\bibnamefont{Prodan}}, \bibinfo{journal}{Phys. Rev. B} \textbf{\bibinfo{volume}{73}}, \bibinfo{pages}{035128} (\bibinfo{year}{2006}), \urlprefix\url{https://link.aps.org/doi/10.1103/PhysRevB.73.035128}.

\bibitem[{\citenamefont{Reuter}(2016)}]{Reuter_2017}
\bibinfo{author}{\bibfnamefont{M.~G.} \bibnamefont{Reuter}}, \bibinfo{journal}{Journal of Physics: Condensed Matter} \textbf{\bibinfo{volume}{29}}, \bibinfo{pages}{053001} (\bibinfo{year}{2016}), \urlprefix\url{https://dx.doi.org/10.1088/1361-648X/29/5/053001}.

\bibitem[{\citenamefont{Bosoni and Sanvito}(2021)}]{Bosoni_2022}
\bibinfo{author}{\bibfnamefont{E.}~\bibnamefont{Bosoni}} \bibnamefont{and} \bibinfo{author}{\bibfnamefont{S.}~\bibnamefont{Sanvito}}, \bibinfo{journal}{Journal of Physics: Condensed Matter} \textbf{\bibinfo{volume}{34}}, \bibinfo{pages}{105501} (\bibinfo{year}{2021}), \urlprefix\url{https://dx.doi.org/10.1088/1361-648X/ac413d}.

\bibitem[{\citenamefont{Kim et~al.}(2022)\citenamefont{Kim, Park, Yoo, Seo, Kim, Park, Kim, Kim, Lee, and Ko}}]{FGT4-2022-localmom}
\bibinfo{author}{\bibfnamefont{S.-R.} \bibnamefont{Kim}}, \bibinfo{author}{\bibfnamefont{I.~K.} \bibnamefont{Park}}, \bibinfo{author}{\bibfnamefont{J.-G.} \bibnamefont{Yoo}}, \bibinfo{author}{\bibfnamefont{J.}~\bibnamefont{Seo}}, \bibinfo{author}{\bibfnamefont{J.-G.} \bibnamefont{Kim}}, \bibinfo{author}{\bibfnamefont{J.-H.} \bibnamefont{Park}}, \bibinfo{author}{\bibfnamefont{J.~S.} \bibnamefont{Kim}}, \bibinfo{author}{\bibfnamefont{K.}~\bibnamefont{Kim}}, \bibinfo{author}{\bibfnamefont{G.}~\bibnamefont{Lee}}, \bibnamefont{and} \bibinfo{author}{\bibfnamefont{K.-T.} \bibnamefont{Ko}}, \bibinfo{journal}{ACS Applied Nano Materials} \textbf{\bibinfo{volume}{5}}, \bibinfo{pages}{10341} (\bibinfo{year}{2022}), \urlprefix\url{https://doi.org/10.1021/acsanm.2c01576}.

\bibitem[{\citenamefont{Sihi~et al.}(2024)}]{antik2}
\bibinfo{author}{\bibfnamefont{A.}~\bibnamefont{Sihi~et al.}} (\bibinfo{year}{2024}), \bibinfo{note}{under preparation}.

\end{thebibliography}

\end{document}